\input harvmac.tex
 \input epsf.tex
 \input amssym

\def\figin{\epsfcheck\figin}\def\figins{\epsfcheck\figins}
\def\epsfcheck{\ifx\epsfbox\UnDeFiNeD
\message{(NO epsf.tex, FIGURES WILL BE IGNORED)}
\gdef\figin##1{\vskip2in}\gdef\figins##1{\hskip.5in}
\else\message{(FIGURES WILL BE INCLUDED)}%
\gdef\figin##1{##1}\gdef\figins##1{##1}\fi}
\def\DefWarn#1{}
\def\figinsert{\goodbreak\midinsert}
\def\ifig#1#2#3{\DefWarn#1\xdef#1{fig.~\the\figno}
\writedef{#1\leftbracket fig.\noexpand~\the\figno} %
\figinsert\figin{\centerline{#3}}\medskip\centerline{\vbox{\baselineskip12pt
\advance\hsize by -1truein\noindent\footnotefont{\bf
Fig.~\the\figno:} #2}}
\bigskip\endinsert\global\advance\figno by1}

\def\unit{\relax{\rm 1\kern-.26em I}}
\def\nada{\relax{\rm 0\kern-.30em l}}
\def\tilde{\widetilde}


\def \la {\langle}
\def \ra {\rangle}
\def \pa {\partial}

\def \eps {\epsilon}


\noblackbox
\def\IL{\relax{\rm I\kern-.18em L}}
\def\IH{\relax{\rm I\kern-.18em H}}
\def\IR{\relax{\rm I\kern-.18em R}}
\def\IC{\relax\hbox{$\inbar\kern-.3em{\rm C}$}}
\def\IZ{\relax\ifmmode\mathchoice
{\hbox{\cmss Z\kern-.4em Z}}{\hbox{\cmss Z\kern-.4em Z}} {\lower.9pt\hbox{\cmsss Z\kern-.4em Z}}
{\lower1.2pt\hbox{\cmsss Z\kern-.4em Z}}\else{\cmss Z\kern-.4em Z}\fi}


\font\manual=manfnt \def\dbend{\lower3.5pt\hbox{\manual\char127}}

\lref\GaberdielUCA{
  M.~R.~Gaberdiel, C.~Peng and I.~G.~Zadeh,
  ``Higgsing the stringy higher spin symmetry,''
[arXiv:1506.02045 [hep-th]].
}

\lref\BuchelSK{
  A.~Buchel, J.~Escobedo, R.~C.~Myers, M.~F.~Paulos, A.~Sinha and M.~Smolkin,
  ``Holographic GB gravity in arbitrary dimensions,''
JHEP {\bf 1003}, 111 (2010).
[arXiv:0911.4257 [hep-th]].
}

\lref\DolanUT{
  F.~A.~Dolan and H.~Osborn,
  ``Conformal four point functions and the operator product expansion,''
Nucl.\ Phys.\ B {\bf 599}, 459 (2001).
[hep-th/0011040].
}

\lref\BelitskyJP{
  A.~V.~Belitsky, J.~Henn, C.~Jarczak, D.~Mueller and E.~Sokatchev,
  ``Anomalous dimensions of leading twist conformal operators,''
Phys.\ Rev.\ D {\bf 77}, 045029 (2008).
[arXiv:0707.2936 [hep-th]].
}

\lref\ManashovFHA{
  A.~N.~Manashov and M.~Strohmaier,
  ``Conformal constraints for anomalous dimensions of leading twist operators,''
[arXiv:1503.04670 [hep-th]].
}

\lref\MaldacenaJN{
  J.~Maldacena and A.~Zhiboedov,
  ``Constraining Conformal Field Theories with A Higher Spin Symmetry,''
J.\ Phys.\ A {\bf 46}, 214011 (2013).
[arXiv:1112.1016 [hep-th]].
}

\lref\MaldacenaSF{
  J.~Maldacena and A.~Zhiboedov,
  ``Constraining conformal field theories with a slightly broken higher spin symmetry,''
Class.\ Quant.\ Grav.\  {\bf 30}, 104003 (2013).
[arXiv:1204.3882 [hep-th]].
}

\lref\FitzpatrickYX{
  A.~L.~Fitzpatrick, J.~Kaplan, D.~Poland and D.~Simmons-Duffin,
  ``The Analytic Bootstrap and AdS Superhorizon Locality,''
JHEP {\bf 1312}, 004 (2013).
[arXiv:1212.3616 [hep-th]].
}

\lref\KomargodskiEK{
  Z.~Komargodski and A.~Zhiboedov,
  ``Convexity and Liberation at Large Spin,''
JHEP {\bf 1311}, 140 (2013).
[arXiv:1212.4103 [hep-th]].
}

\lref\AldayZY{
  L.~F.~Alday, B.~Eden, G.~P.~Korchemsky, J.~Maldacena and E.~Sokatchev,
  ``From correlation functions to Wilson loops,''
JHEP {\bf 1109}, 123 (2011).
[arXiv:1007.3243 [hep-th]].
}

\lref\BelavinVU{
  A.~A.~Belavin, A.~M.~Polyakov and A.~B.~Zamolodchikov,
  ``Infinite Conformal Symmetry in Two-Dimensional Quantum Field Theory,''
Nucl.\ Phys.\ B {\bf 241}, 333 (1984)..
}

\lref\joao{
  J.~Penedones,
unpublished.
}

\lref\PolyakovHA{
  A.~M.~Polyakov,
  ``Nonhamiltonian Approach To The Quantum Field Theory At Small Distances,''
Zh. Eksp. Teor. Fiz.\ {\bf 66}, 23 (1974)
}

\lref\FerraraYT{
  S.~Ferrara, A.~F.~Grillo and R.~Gatto,
  ``Tensor representations of conformal algebra and conformally covariant operator product expansion,''
Annals Phys.\  {\bf 76}, 161 (1973).
}

\lref\RattazziPE{
  R.~Rattazzi, V.~S.~Rychkov, E.~Tonni and A.~Vichi,
  ``Bounding scalar operator dimensions in 4D CFT,''
JHEP {\bf 0812}, 031 (2008).
[arXiv:0807.0004 [hep-th]].
}

\lref\RychkovIJ{
  V.~S.~Rychkov and A.~Vichi,
  ``Universal Constraints on Conformal Operator Dimensions,''
Phys.\ Rev.\ D {\bf 80}, 045006 (2009).
[arXiv:0905.2211 [hep-th]].
}

\lref\ElShowkHT{
  S.~El-Showk, M.~F.~Paulos, D.~Poland, S.~Rychkov, D.~Simmons-Duffin and A.~Vichi,
  ``Solving the 3D Ising Model with the Conformal Bootstrap,''
Phys.\ Rev.\ D {\bf 86}, 025022 (2012).
[arXiv:1203.6064 [hep-th]].
}

\lref\AldayCWA{
  L.~F.~Alday and A.~Bissi,
  ``Higher-spin correlators,''
JHEP {\bf 1310}, 202 (2013).
[arXiv:1305.4604 [hep-th]].
}

\lref\KlebanovJA{
  I.~R.~Klebanov and A.~M.~Polyakov,
  ``AdS dual of the critical O(N) vector model,''
Phys.\ Lett.\ B {\bf 550}, 213 (2002).
[hep-th/0210114].
}

\lref\VasilievBA{
  M.~A.~Vasiliev,
In *Shifman, M.A. (ed.): The many faces of the superworld* 533-610.
[hep-th/9910096].
}

\lref\GiombiWH{
  S.~Giombi and X.~Yin,
JHEP {\bf 1009}, 115 (2010).
[arXiv:0912.3462 [hep-th]].
}

\lref\CallanPU{
  C.~G.~Callan, Jr. and D.~J.~Gross,
  ``Bjorken scaling in quantum field theory,''
Phys.\ Rev.\ D {\bf 8}, 4383 (1973)..
}

\lref\AldayCWA{
  L.~F.~Alday and A.~Bissi,
  ``Higher-spin correlators,''
JHEP {\bf 1310}, 202 (2013).
[arXiv:1305.4604 [hep-th]].
}

\lref\CardyYY{
  J.~L.~Cardy,
  ``Conformal Invariance and the Yang-lee Edge Singularity in Two-dimensions,''
Phys.\ Rev.\ Lett.\  {\bf 54}, 1354 (1985)..
}

\lref\AldayEYA{
  L.~F.~Alday, A.~Bissi and T.~Lukowski,
  ``Large spin systematics in CFT,''
[arXiv:1502.07707 [hep-th]].
}

\lref\DolanDV{
  F.~A.~Dolan and H.~Osborn,
  ``Conformal Partial Waves: Further Mathematical Results,''
[arXiv:1108.6194 [hep-th]].
}

\lref\NachtmannMR{
  O.~Nachtmann,
  ``Positivity constraints for anomalous dimensions,''
Nucl.\ Phys.\ B {\bf 63}, 237 (1973)..
}

\lref\GiddingsGJ{
  S.~B.~Giddings and R.~A.~Porto,
  ``The Gravitational S-matrix,''
Phys.\ Rev.\ D {\bf 81}, 025002 (2010).
[arXiv:0908.0004 [hep-th]].
}

\lref\NachtmannMR{
  O.~Nachtmann,
  ``Positivity constraints for anomalous dimensions,''
Nucl.\ Phys.\ B {\bf 63}, 237 (1973)..
}

\lref\HeemskerkPN{
  I.~Heemskerk, J.~Penedones, J.~Polchinski and J.~Sully,
  ``Holography from Conformal Field Theory,''
JHEP {\bf 0910}, 079 (2009).
[arXiv:0907.0151 [hep-th]].
}

\lref\FitzpatrickZM{
  A.~L.~Fitzpatrick, E.~Katz, D.~Poland and D.~Simmons-Duffin,
  ``Effective Conformal Theory and the Flat-Space Limit of AdS,''
JHEP {\bf 1107}, 023 (2011).
[arXiv:1007.2412 [hep-th]].
}

\lref\NachtmannMR{
  O.~Nachtmann,
  ``Positivity constraints for anomalous dimensions,''
Nucl.\ Phys.\ B {\bf 63}, 237 (1973).
}

\lref\BanksBJ{
  T.~Banks and G.~Festuccia,
  ``The Regge Limit for Green Functions in Conformal Field Theory,''
JHEP {\bf 1006}, 105 (2010).
[arXiv:0910.2746 [hep-th]].
}

\lref\Eden{
  R.~J.~Eden, P.~V.~Landshoff, D.~I.~Olive and J.~C.~Polkinghorne,
  ``The analytic S matrix''
}

\lref\PolyakovGS{
  A.~M.~Polyakov,
  ``Nonhamiltonian approach to conformal quantum field theory,''
Zh.\ Eksp.\ Teor.\ Fiz.\  {\bf 66}, 23 (1974).
}

\lref\HofmanAR{
  D.~M.~Hofman and J.~Maldacena,
  ``Conformal collider physics: Energy and charge correlations,''
JHEP {\bf 0805}, 012 (2008).
[arXiv:0803.1467 [hep-th]].
}

\lref\AldayMF{
  L.~F.~Alday and J.~M.~Maldacena,
  ``Comments on operators with large spin,''
JHEP {\bf 0711}, 019 (2007).
[arXiv:0708.0672 [hep-th]].
}

\lref\AldayZY{
  L.~F.~Alday, B.~Eden, G.~P.~Korchemsky, J.~Maldacena and E.~Sokatchev,
  ``From correlation functions to Wilson loops,''
JHEP {\bf 1109}, 123 (2011).
[arXiv:1007.3243 [hep-th]].
}

\lref\DolanDV{
  F.~A.~Dolan and H.~Osborn,
  ``Conformal Partial Waves: Further Mathematical Results,''
[arXiv:1108.6194 [hep-th]].
}

\lref\ElShowkHT{
  S.~El-Showk, M.~F.~Paulos, D.~Poland, S.~Rychkov, D.~Simmons-Duffin and A.~Vichi,
  ``Solving the 3D Ising Model with the Conformal Bootstrap,''
[arXiv:1203.6064 [hep-th]].
}

\lref\WilsonJJ{
  K.~G.~Wilson and J.~B.~Kogut,
  ``The Renormalization group and the epsilon expansion,''
Phys.\ Rept.\  {\bf 12}, 75 (1974)..
}

\lref\LangZW{
  K.~Lang and W.~Ruhl,
  ``The Critical O(N) sigma model at dimensions 2 < d < 4: Fusion coefficients and anomalous dimensions,''
Nucl.\ Phys.\ B {\bf 400}, 597 (1993)..
}

\lref\HoffmannDX{
  L.~Hoffmann, L.~Mesref and W.~Ruhl,
  ``Conformal partial wave analysis of AdS amplitudes for dilaton axion four point functions,''
Nucl.\ Phys.\ B {\bf 608}, 177 (2001).
[hep-th/0012153].
}

\lref\ElShowkHT{
  S.~El-Showk, M.~F.~Paulos, D.~Poland, S.~Rychkov, D.~Simmons-Duffin and A.~Vichi,
  ``Solving the 3D Ising Model with the Conformal Bootstrap,''
[arXiv:1203.6064 [hep-th]].
}

\lref\GrossUN{
 D. Gross, {\it unpublished} 
}

\lref\CallanPU{
  C.~G.~Callan, Jr. and D.~J.~Gross,
  ``Bjorken scaling in quantum field theory,''
Phys.\ Rev.\ D {\bf 8}, 4383 (1973)..
}

\lref\MackJE{
  G.~Mack,
 ``All Unitary Ray Representations of the Conformal Group SU(2,2) with Positive Energy,''
Commun.\ Math.\ Phys.\  {\bf 55}, 1 (1977)..
}

\lref\GrinsteinQK{
  B.~Grinstein, K.~A.~Intriligator and I.~Z.~Rothstein,
  ``Comments on Unparticles,''
Phys.\ Lett.\ B {\bf 662}, 367 (2008).
[arXiv:0801.1140 [hep-ph]].
}

\lref\RattazziPE{
  R.~Rattazzi, V.~S.~Rychkov, E.~Tonni and A.~Vichi,
  ``Bounding scalar operator dimensions in 4D CFT,''
JHEP {\bf 0812}, 031 (2008).
[arXiv:0807.0004 [hep-th]].
}

\lref\CostaCB{
  M.~S.~Costa, V.~Goncalves and J.~Penedones,
  ``Conformal Regge theory,''
[arXiv:1209.4355 [hep-th]].
}

\lref\PappadopuloJK{
  D.~Pappadopulo, S.~Rychkov, J.~Espin and R.~Rattazzi,
  ``OPE Convergence in Conformal Field Theory,''
[arXiv:1208.6449 [hep-th]].
}

\lref\AdamsSV{
  A.~Adams, N.~Arkani-Hamed, S.~Dubovsky, A.~Nicolis and R.~Rattazzi,
  ``Causality, analyticity and an IR obstruction to UV completion,''
JHEP {\bf 0610}, 014 (2006).
[hep-th/0602178].
}

\lref\FreyhultMY{
  L.~Freyhult and S.~Zieme,
  ``The virtual scaling function of AdS/CFT,''
Phys.\ Rev.\ D {\bf 79}, 105009 (2009).
[arXiv:0901.2749 [hep-th]].
}

\lref\PelissettoEK{
  A.~Pelissetto and E.~Vicari,
  ``Critical phenomena and renormalization group theory,''
Phys.\ Rept.\  {\bf 368}, 549 (2002).
[cond-mat/0012164].
}

\lref\VasilievBA{
  M.~A.~Vasiliev,
  ``Higher spin gauge theories: Star product and AdS space,''
In *Shifman, M.A. (ed.): The many faces of the superworld* 533-610.
[hep-th/9910096].
}

\lref\MaldacenaRE{
  J.~M.~Maldacena,
  ``The Large N limit of superconformal field theories and supergravity,''
Adv.\ Theor.\ Math.\ Phys.\  {\bf 2}, 231 (1998).
[hep-th/9711200].
}

\lref\GubserBC{
  S.~S.~Gubser, I.~R.~Klebanov and A.~M.~Polyakov,
  ``Gauge theory correlators from noncritical string theory,''
Phys.\ Lett.\ B {\bf 428}, 105 (1998).
[hep-th/9802109].
}

\lref\WittenQJ{
  E.~Witten,
  ``Anti-de Sitter space and holography,''
Adv.\ Theor.\ Math.\ Phys.\  {\bf 2}, 253 (1998).
[hep-th/9802150].
}

\lref\SundborgWP{
  B.~Sundborg,
  ``Stringy gravity, interacting tensionless strings and massless higher spins,''
Nucl.\ Phys.\ Proc.\ Suppl.\  {\bf 102}, 113 (2001).
[hep-th/0103247].
}

\lref\DolanUT{
  F.~A.~Dolan and H.~Osborn,
  ``Conformal four point functions and the operator product expansion,''
Nucl.\ Phys.\ B {\bf 599}, 459 (2001).
[hep-th/0011040].
}

\lref\CornalbaXK{
  L.~Cornalba, M.~S.~Costa, J.~Penedones and R.~Schiappa,
  ``Eikonal Approximation in AdS/CFT: From Shock Waves to Four-Point Functions,''
JHEP {\bf 0708}, 019 (2007).
[hep-th/0611122].
}

\lref\CornalbaXM{
  L.~Cornalba, M.~S.~Costa, J.~Penedones and R.~Schiappa,
  ``Eikonal Approximation in AdS/CFT: Conformal Partial Waves and Finite N Four-Point Functions,''
Nucl.\ Phys.\ B {\bf 767}, 327 (2007).
[hep-th/0611123].
}

\lref\CornalbaZB{
  L.~Cornalba, M.~S.~Costa and J.~Penedones,
  ``Eikonal approximation in AdS/CFT: Resumming the gravitational loop expansion,''
JHEP {\bf 0709}, 037 (2007).
[arXiv:0707.0120 [hep-th]].
}

\lref\DerkachovPH{
  S.~E.~Derkachov and A.~N.~Manashov,
  ``Generic scaling relation in the scalar phi**4 model,''
J.\ Phys.\ A {\bf 29}, 8011 (1996).
[hep-th/9604173].
}

\lref\KlebanovJA{
  I.~R.~Klebanov and A.~M.~Polyakov,
  ``AdS dual of the critical O(N) vector model,''
Phys.\ Lett.\ B {\bf 550}, 213 (2002).
[hep-th/0210114].
}

\lref\SezginRT{
  E.~Sezgin and P.~Sundell,
  ``Massless higher spins and holography,''
Nucl.\ Phys.\ B {\bf 644}, 303 (2002), [Erratum-ibid.\ B {\bf 660}, 403 (2003)].
[hep-th/0205131].
}
\lref\EpsteinBG{
  H.~Epstein, V.~Glaser and A.~Martin,
  ``Polynomial behaviour of scattering amplitudes at fixed momentum transfer in theories with local observables,''
Commun.\ Math.\ Phys.\  {\bf 13}, 257 (1969)..
}

\lref\FortinCQ{
  J.~-F.~Fortin, B.~Grinstein and A.~Stergiou,
  ``Scale without Conformal Invariance in Four Dimensions,''
[arXiv:1206.2921 [hep-th]].
}

\lref\FortinHN{
  J.~-F.~Fortin, B.~Grinstein and A.~Stergiou,
  ``A generalized c-theorem and the consistency of scale without conformal invariance,''
[arXiv:1208.3674 [hep-th]].
}

\lref\FortinHC{
  J.~-F.~Fortin, B.~Grinstein, C.~W.~Murphy and A.~Stergiou,
  ``On Limit Cycles in Supersymmetric Theories,''
[arXiv:1210.2718 [hep-th]].
}

\lref\NakayamaED{
  Y.~Nakayama,
  ``Is boundary conformal in CFT?,''
[arXiv:1210.6439 [hep-th]].
}

\lref\NakayamaND{
  Y.~Nakayama,
  ``Supercurrent, Supervirial and Superimprovement,''
[arXiv:1208.4726 [hep-th]].
}

\lref\AntoniadisGN{
  I.~Antoniadis and M.~Buican,
  ``On R-symmetric Fixed Points and Superconformality,''
Phys.\ Rev.\ D {\bf 83}, 105011 (2011).
[arXiv:1102.2294 [hep-th]].
}

\lref\LutyWW{
  M.~A.~Luty, J.~Polchinski and R.~Rattazzi,
  ``The $a$-theorem and the Asymptotics of 4D Quantum Field Theory,''
[arXiv:1204.5221 [hep-th]].
}

\lref\GiombiWH{
  S.~Giombi and X.~Yin,
  ``Higher Spin Gauge Theory and Holography: The Three-Point Functions,''
JHEP {\bf 1009}, 115 (2010).
[arXiv:0912.3462 [hep-th]].
}

\lref\KomargodskiVJ{
  Z.~Komargodski and A.~Schwimmer,
  ``On Renormalization Group Flows in Four Dimensions,''
JHEP {\bf 1112}, 099 (2011).
[arXiv:1107.3987 [hep-th]].
}

\lref\KomargodskiXV{
  Z.~Komargodski,
  ``The Constraints of Conformal Symmetry on RG Flows,''
JHEP {\bf 1207}, 069 (2012).
[arXiv:1112.4538 [hep-th]].
}

\lref\KehreinIA{
  S.~K.~Kehrein,
  ``The Spectrum of critical exponents in phi**2 in two-dimensions theory in D = (4-epsilon)-dimensions: Resolution of degeneracies and hierarchical structures,''
Nucl.\ Phys.\ B {\bf 453}, 777 (1995).
[hep-th/9507044].
}

\lref\BraunRP{
  V.~M.~Braun, G.~P.~Korchemsky and D.~Mueller,
  ``The Uses of conformal symmetry in QCD,''
Prog.\ Part.\ Nucl.\ Phys.\  {\bf 51}, 311 (2003).
[hep-ph/0306057].
}

\lref\HeemskerkPN{
  I.~Heemskerk, J.~Penedones, J.~Polchinski and J.~Sully,
  ``Holography from Conformal Field Theory,''
JHEP {\bf 0910}, 079 (2009).
[arXiv:0907.0151 [hep-th]].
}

\lref\RuhlBW{
  W.~Ruhl,
  ``The Goldstone fields of interacting higher spin field theory on AdS(4),''
[hep-th/0607197].
}

\lref\AdamsSV{
  A.~Adams, N.~Arkani-Hamed, S.~Dubovsky, A.~Nicolis and R.~Rattazzi,
  ``Causality, analyticity and an IR obstruction to UV completion,''
JHEP {\bf 0610}, 014 (2006).
[hep-th/0602178].
}

\lref\PhamCR{
  T.~N.~Pham and T.~N.~Truong,
  ``Evaluation Of The Derivative Quartic Terms Of The Meson Chiral Lagrangian From Forward Dispersion Relation,''
Phys.\ Rev.\ D {\bf 31}, 3027 (1985)..
}

\lref\FGG{
  S.~Ferrara, A.~F.~Grillo and R.~Gatto,
  ``Tensor representations of conformal algebra and conformally covariant operator product expansion,''
Annals Phys.\  {\bf 76}, 161 (1973).
}

\lref\BPZ{
  A.~A.~Belavin, A.~M.~Polyakov and A.~B.~Zamolodchikov,
  ``Infinite Conformal Symmetry in Two-Dimensional Quantum Field Theory,''
Nucl.\ Phys.\ B {\bf 241}, 333 (1984).
}

\lref\HeslopDU{
  P.~J.~Heslop,
  ``Aspects of superconformal field theories in six dimensions,''
JHEP {\bf 0407}, 056 (2004).
[hep-th/0405245].
}

\lref\DolanTT{
  F.~A.~Dolan and H.~Osborn,
  ``Superconformal symmetry, correlation functions and the operator product expansion,''
Nucl.\ Phys.\ B {\bf 629}, 3 (2002).
[hep-th/0112251].
}

\lref\PappadopuloJK{
  D.~Pappadopulo, S.~Rychkov, J.~Espin and R.~Rattazzi,
  ``OPE Convergence in Conformal Field Theory,''
[arXiv:1208.6449 [hep-th]].
}

\lref\FerraraEJ{
  S.~Ferrara, C.~Fronsdal and A.~Zaffaroni,
  ``On N=8 supergravity on AdS(5) and N=4 superconformal Yang-Mills theory,''
Nucl.\ Phys.\ B {\bf 532}, 153 (1998).
[hep-th/9802203].
}

\lref\DidenkoTV{
  V.~E.~Didenko and E.~D.~Skvortsov,
  ``Exact higher-spin symmetry in CFT: all correlators in unbroken Vasiliev theory,''
[arXiv:1210.7963 [hep-th]].
}

\lref\MaldacenaSF{
  J.~Maldacena and A.~Zhiboedov,
  ``Constraining conformal field theories with a slightly broken higher spin symmetry,''
[arXiv:1204.3882 [hep-th]].
}

\lref\MaldacenaJN{
  J.~Maldacena and A.~Zhiboedov,
  ``Constraining Conformal Field Theories with A Higher Spin Symmetry,''
[arXiv:1112.1016 [hep-th]].
}

\lref\BriganteGZ{
  M.~Brigante, H.~Liu, R.~C.~Myers, S.~Shenker and S.~Yaida,
  ``The Viscosity Bound and Causality Violation,''
Phys.\ Rev.\ Lett.\  {\bf 100}, 191601 (2008).
[arXiv:0802.3318 [hep-th]].
}

\lref\GirardelloPP{
  L.~Girardello, M.~Porrati and A.~Zaffaroni,
  ``3-D interacting CFTs and generalized Higgs phenomenon in higher spin theories on AdS,''
Phys.\ Lett.\ B {\bf 561}, 289 (2003).
[hep-th/0212181].
}

\lref\HeemskerkPN{
  I.~Heemskerk, J.~Penedones, J.~Polchinski and J.~Sully,
  ``Holography from Conformal Field Theory,''
JHEP {\bf 0910}, 079 (2009).
[arXiv:0907.0151 [hep-th]].
}

\lref\CornalbaZB{
  L.~Cornalba, M.~S.~Costa and J.~Penedones,
  ``Eikonal approximation in AdS/CFT: Resumming the gravitational loop expansion,''
JHEP {\bf 0709}, 037 (2007).
[arXiv:0707.0120 [hep-th]].
}

\lref\DolanDV{
  F.~A.~Dolan and H.~Osborn,
  ``Conformal Partial Waves: Further Mathematical Results,''
[arXiv:1108.6194 [hep-th]].
}

\lref\DolanUT{
  F.~A.~Dolan and H.~Osborn,
  ``Conformal four point functions and the operator product expansion,''
Nucl.\ Phys.\ B {\bf 599}, 459 (2001).
[hep-th/0011040].
}

\lref\GiombiKC{
  S.~Giombi, S.~Minwalla, S.~Prakash, S.~P.~Trivedi, S.~R.~Wadia and X.~Yin,
  ``Chern-Simons Theory with Vector Fermion Matter,''
Eur.\ Phys.\ J.\ C {\bf 72}, 2112 (2012).
[arXiv:1110.4386 [hep-th]].
}

\lref\AharonyJZ{
  O.~Aharony, G.~Gur-Ari and R.~Yacoby,
  ``d=3 Bosonic Vector Models Coupled to Chern-Simons Gauge Theories,''
JHEP {\bf 1203}, 037 (2012).
[arXiv:1110.4382 [hep-th]].
}

\lref\KovtunKW{
  P.~Kovtun and A.~Ritz,
  ``Black holes and universality classes of critical points,''
Phys.\ Rev.\ Lett.\  {\bf 100}, 171606 (2008).
[arXiv:0801.2785 [hep-th]].
}

\lref\FroissartUX{
  M.~Froissart,
  ``Asymptotic behavior and subtractions in the Mandelstam representation,''
Phys.\ Rev.\  {\bf 123}, 1053 (1961)..
}

\lref\MartinRT{
  A.~Martin,
  ``Unitarity and high-energy behavior of scattering amplitudes,''
Phys.\ Rev.\  {\bf 129}, 1432 (1963)..
}

\lref\FrankfurtMC{
  L.~Frankfurt, M.~Strikman and C.~Weiss,
 ``Small-x physics: From HERA to LHC and beyond,''
Ann.\ Rev.\ Nucl.\ Part.\ Sci.\  {\bf 55}, 403 (2005).
[hep-ph/0507286].
}

\lref\PolchinskiJW{
  J.~Polchinski and M.~J.~Strassler,
  ``Deep inelastic scattering and gauge / string duality,''
JHEP {\bf 0305}, 012 (2003).
[hep-th/0209211].
}

\lref\BarnesBM{
  E.~Barnes, E.~Gorbatov, K.~A.~Intriligator, M.~Sudano and J.~Wright,
  ``The Exact superconformal R-symmetry minimizes tau(RR),''
Nucl.\ Phys.\ B {\bf 730}, 210 (2005).
[hep-th/0507137].
}

\lref\PolandWG{
  D.~Poland and D.~Simmons-Duffin,
  ``Bounds on 4D Conformal and Superconformal Field Theories,''
JHEP {\bf 1105}, 017 (2011).
[arXiv:1009.2087 [hep-th]].
}

\lref\LiendoHY{
  P.~Liendo, L.~Rastelli and B.~C.~van Rees,
  ``The Bootstrap Program for Boundary $CFT_d$,''
[arXiv:1210.4258 [hep-th]].
}

\lref\PolyakovXD{
 A.~M.~Polyakov,
 ``Conformal symmetry of critical fluctuations,''
JETP Lett.\  {\bf 12}, 381 (1970), [Pisma Zh.\ Eksp.\ Teor.\ Fiz.\  {\bf 12}, 538 (1970)].
}

\lref\FitzpatrickDM{
  A.~L.~Fitzpatrick and J.~Kaplan,
  ``Unitarity and the Holographic S-Matrix,''
JHEP {\bf 1210}, 032 (2012).
[arXiv:1112.4845 [hep-th]].
}

\lref\CostaDW{
  M.~S.~Costa, J.~Penedones, D.~Poland and S.~Rychkov,
  ``Spinning Conformal Blocks,''
JHEP {\bf 1111}, 154 (2011).
[arXiv:1109.6321 [hep-th]].
}

\lref\OsbornVT{
  H.~Osborn,
  ``Conformal Blocks for Arbitrary Spins in Two Dimensions,''
Phys.\ Lett.\ B {\bf 718}, 169 (2012).
[arXiv:1205.1941 [hep-th]].
}

\lref\LangGE{
  K.~Lang and W.~Ruhl,
  ``Critical O(N) vector nonlinear sigma models: A Resume of their field structure,''
[hep-th/9311046].
}

\lref\FitzpatrickYX{
  A.~L.~Fitzpatrick, J.~Kaplan, D.~Poland and D.~Simmons-Duffin,
  ``The Analytic Bootstrap and AdS Superhorizon Locality,''
[arXiv:1212.3616 [hep-th]].
}

\lref\ArutyunovFH{
  G.~Arutyunov, F.~A.~Dolan, H.~Osborn and E.~Sokatchev,
  ``Correlation functions and massive Kaluza-Klein modes in the AdS / CFT correspondence,''
Nucl.\ Phys.\ B {\bf 665}, 273 (2003).
[hep-th/0212116].
}

\lref\BianchiCM{
  M.~Bianchi, S.~Kovacs, G.~Rossi and Y.~S.~Stanev,
  ``Properties of the Konishi multiplet in N=4 SYM theory,''
JHEP {\bf 0105}, 042 (2001).
[hep-th/0104016].
}

\lref\LangKP{
  K.~Lang and W.~Ruhl,
  ``The Critical O(N) sigma model at dimension 2 < d < 4 and order 1/n**2: Operator product expansions and renormalization,''
Nucl.\ Phys.\ B {\bf 377}, 371 (1992)..
}

\lref\DolanIY{
  F.~A.~Dolan and H.~Osborn,
  ``Conformal partial wave expansions for N=4 chiral four point functions,''
Annals Phys.\  {\bf 321}, 581 (2006).
[hep-th/0412335].
}

\lref\PappadopuloJK{
  D.~Pappadopulo, S.~Rychkov, J.~Espin and R.~Rattazzi,
  ``OPE Convergence in Conformal Field Theory,''
Phys.\ Rev.\ D {\bf 86}, 105043 (2012).
[arXiv:1208.6449 [hep-th]].
}

\lref\SimmonsDuffinQMA{
  D.~Simmons-Duffin,
  ``A Semidefinite Program Solver for the Conformal Bootstrap,''
[arXiv:1502.02033 [hep-th]].
}

\lref\ElShowkHT{
  S.~El-Showk, M.~F.~Paulos, D.~Poland, S.~Rychkov, D.~Simmons-Duffin and A.~Vichi,
  ``Solving the 3D Ising Model with the Conformal Bootstrap,''
Phys.\ Rev.\ D {\bf 86}, 025022 (2012).
[arXiv:1203.6064 [hep-th]].
}

\lref\ElShowkDWA{
  S.~El-Showk, M.~F.~Paulos, D.~Poland, S.~Rychkov, D.~Simmons-Duffin and A.~Vichi,
  ``Solving the 3d Ising Model with the Conformal Bootstrap II. c-Minimization and Precise Critical Exponents,''
J.\ Stat.\ Phys.\  {\bf 157}, 869 (2014).
[arXiv:1403.4545 [hep-th]].
}

\lref\GrossUE{
  D.~J.~Gross,
  ``High-Energy Symmetries of String Theory,''
Phys.\ Rev.\ Lett.\  {\bf 60}, 1229 (1988)..
}

\lref\GaberdielCHA{
  M.~R.~Gaberdiel and R.~Gopakumar,
  ``Higher Spins \& Strings,''
JHEP {\bf 1411}, 044 (2014).
[arXiv:1406.6103 [hep-th]].
}

\lref\VasilievEV{
  M.~A.~Vasiliev,
  ``Nonlinear equations for symmetric massless higher spin fields in (A)dS(d),''
Phys.\ Lett.\ B {\bf 567}, 139 (2003).
[hep-th/0304049].
}

\lref\StanevQRA{
  Y.~S.~Stanev,
  ``Constraining conformal field theory with higher spin symmetry in four dimensions,''
Nucl.\ Phys.\ B {\bf 876}, 651 (2013).
[arXiv:1307.5209 [hep-th]].
}

\lref\AlbaYDA{
  V.~Alba and K.~Diab,
  ``Constraining conformal field theories with a higher spin symmetry in d=4,''
[arXiv:1307.8092 [hep-th]].
}

\lref\BoulangerZZA{
  N.~Boulanger, D.~Ponomarev, E.~D.~Skvortsov and M.~Taronna,
  ``On the uniqueness of higher-spin symmetries in AdS and CFT,''
Int.\ J.\ Mod.\ Phys.\ A {\bf 28}, 1350162 (2013).
[arXiv:1305.5180 [hep-th]].
}

\lref\PolyakovXD{
  A.~M.~Polyakov,
  ``Conformal symmetry of critical fluctuations,''
JETP Lett.\  {\bf 12}, 381 (1970), [Pisma Zh.\ Eksp.\ Teor.\ Fiz.\  {\bf 12}, 538 (1970)]..
}

\lref\ElShowkDWA{
  S.~El-Showk, M.~F.~Paulos, D.~Poland, S.~Rychkov, D.~Simmons-Duffin and A.~Vichi,
  ``Solving the 3d Ising Model with the Conformal Bootstrap II. c-Minimization and Precise Critical Exponents,''
J.\ Stat.\ Phys.\  {\bf 157}, 869 (2014).
[arXiv:1403.4545 [hep-th]].
}

\lref\EdenBK{
  B.~Eden, A.~C.~Petkou, C.~Schubert and E.~Sokatchev,
  ``Partial nonrenormalization of the stress tensor four point function in N=4 SYM and AdS / CFT,''
Nucl.\ Phys.\ B {\bf 607}, 191 (2001).
[hep-th/0009106].
}

\lref\EdenWE{
  B.~Eden, P.~Heslop, G.~P.~Korchemsky and E.~Sokatchev,
  ``Hidden symmetry of four-point correlation functions and amplitudes in N=4 SYM,''
Nucl.\ Phys.\ B {\bf 862}, 193 (2012).
[arXiv:1108.3557 [hep-th]].
}

\lref\EdenMV{
  B.~Eden, C.~Schubert and E.~Sokatchev,
  ``Three loop four point correlator in N=4 SYM,''
Phys.\ Lett.\ B {\bf 482}, 309 (2000).
[hep-th/0003096].
}

\lref\BianchiHN{
  M.~Bianchi, S.~Kovacs, G.~Rossi and Y.~S.~Stanev,
  ``Anomalous dimensions in N=4 SYM theory at order g**4,''
Nucl.\ Phys.\ B {\bf 584}, 216 (2000).
[hep-th/0003203].
}

\lref\DrummondNDA{
  J.~Drummond, C.~Duhr, B.~Eden, P.~Heslop, J.~Pennington and V.~A.~Smirnov,
  ``Leading singularities and off-shell conformal integrals,''
JHEP {\bf 1308}, 133 (2013).
[arXiv:1303.6909, arXiv:1303.6909 [hep-th]].
}

\lref\HeemskerkPN{
  I.~Heemskerk, J.~Penedones, J.~Polchinski and J.~Sully,
  ``Holography from Conformal Field Theory,''
JHEP {\bf 0910}, 079 (2009).
[arXiv:0907.0151 [hep-th]].
}

\lref\ElShowkHT{
  S.~El-Showk, M.~F.~Paulos, D.~Poland, S.~Rychkov, D.~Simmons-Duffin and A.~Vichi,
  ``Solving the 3D Ising Model with the Conformal Bootstrap,''
Phys.\ Rev.\ D {\bf 86}, 025022 (2012).
[arXiv:1203.6064 [hep-th]].
}

\lref\ArutyunovPY{
  G.~Arutyunov and S.~Frolov,
  ``Four point functions of lowest weight CPOs in N=4 SYM(4) in supergravity approximation,''
Phys.\ Rev.\ D {\bf 62}, 064016 (2000).
[hep-th/0002170].
}

\lref\DolanTT{
  F.~A.~Dolan and H.~Osborn,
  ``Superconformal symmetry, correlation functions and the operator product expansion,''
Nucl.\ Phys.\ B {\bf 629}, 3 (2002).
[hep-th/0112251].
}

\lref\DHokerPJ{
  E.~D'Hoker, D.~Z.~Freedman, S.~D.~Mathur, A.~Matusis and L.~Rastelli,
  ``Graviton exchange and complete four point functions in the AdS / CFT correspondence,''
Nucl.\ Phys.\ B {\bf 562}, 353 (1999).
[hep-th/9903196].
}

\lref\AldayTSA{
  L.~F.~Alday, A.~Bissi and T.~Lukowski,
  ``Lessons from crossing symmetry at large N,''
[arXiv:1410.4717 [hep-th]].
}

\lref\BianchiNK{
  M.~Bianchi, M.~B.~Green, S.~Kovacs and G.~Rossi,
  ``Instantons in supersymmetric Yang-Mills and D instantons in IIB superstring theory,''
JHEP {\bf 9808}, 013 (1998).
[hep-th/9807033].
}

\lref\DoreyPD{
  N.~Dorey, T.~J.~Hollowood, V.~V.~Khoze, M.~P.~Mattis and S.~Vandoren,
  ``Multi-instanton calculus and the AdS / CFT correspondence in N=4 superconformal field theory,''
Nucl.\ Phys.\ B {\bf 552}, 88 (1999).
[hep-th/9901128].
}

\lref\CampostriniAT{
  M.~Campostrini, A.~Pelissetto, P.~Rossi and E.~Vicari,
  ``Improved high temperature expansion and critical equation of state of three-dimensional Ising - like systems,''
Phys.\ Rev.\ E {\bf 60}, 3526 (1999).
[cond-mat/9905078].
}

\lref\KavirajCXA{
  A.~Kaviraj, K.~Sen and A.~Sinha,
  ``Analytic bootstrap at large spin,''
[arXiv:1502.01437 [hep-th]].
}

\lref\KavirajXSA{
  A.~Kaviraj, K.~Sen and A.~Sinha,
  ``Universal anomalous dimensions at large spin and large twist,''
[arXiv:1504.00772 [hep-th]].
}

\lref\FeiYJA{
  L.~Fei, S.~Giombi and I.~R.~Klebanov,
  ``Critical $O(N)$ models in $6-\epsilon$ dimensions,''
Phys.\ Rev.\ D {\bf 90}, no. 2, 025018 (2014).
[arXiv:1404.1094 [hep-th]].
}

\lref\ChesterGQA{
  S.~M.~Chester, S.~S.~Pufu and R.~Yacoby,
  ``Bootstrapping $O(N)$ vector models in 4 $< d <$ 6,''
Phys.\ Rev.\ D {\bf 91}, no. 8, 086014 (2015).
[arXiv:1412.7746 [hep-th]].
}

\lref\PenedonesUE{
  J.~Penedones,
  ``Writing CFT correlation functions as AdS scattering amplitudes,''
JHEP {\bf 1103}, 025 (2011).
[arXiv:1011.1485 [hep-th]].
}

\Title{
\vbox{\baselineskip6pt
}}
{\vbox{\centerline{Conformal Bootstrap With}\vskip .30cm
 \centerline{Slightly Broken Higher Spin Symmetry}
}}

\centerline{Luis F. Alday$^{u}$ and Alexander Zhiboedov$^{v}$  }
\bigskip
\centerline{\it $^{u}$ Mathematical Institute, University of Oxford,} 
\centerline{\it  Andrew Wiles Building, Radcliffe Observatory Quarter,}
\centerline{\it Woodstock Road, Oxford, OX2 6GG, UK}
\centerline{\it $^{v}$ Center for the Fundamental Laws of Nature,}
\centerline{\it Harvard University, Cambridge, MA 02138 USA}

\vskip .5in \noindent
We consider conformal field theories with slightly broken higher spin symmetry in arbitrary spacetime dimensions. We analyze the crossing equation in the double light-cone limit and solve for the anomalous dimensions of higher spin currents $\gamma_s$ with large spin $s$. The result depends on the symmetries and the spectrum of the unperturbed conformal field theory. We reproduce all known results and make further predictions. In particular we make a prediction for the anomalous dimensions of higher spin currents in the 3d Ising model.

\Date{ }

\listtoc\writetoc
\vskip .5in \noindent


\newsec{Introduction}

In this paper we consider higher dimensional conformal field theories CFT$_{d>2}$ with slightly broken higher spin symmetry or, in other words, theories which have higher spin currents with small anomalous dimension $\gamma_s \ll 1$. A typical example of this type are weakly coupled theories.
Weakly coupled theories are usually analyzed using standard perturbative methods that involve computations of Feynman or Feynman-like diagrams. Our approach is to use methods of analytic bootstrap \refs{\FitzpatrickYX\KomargodskiEK-\AldayCWA} combined with the known structure of the free theory correlators. Indeed, when higher spin symmetry is present CFT$_{d>2}$ are believed to be free and thus exactly solvable \refs{\MaldacenaJN\StanevQRA\AlbaYDA-\BoulangerZZA}.
Therefore a natural question is how to use this symmetry to constrain the dynamics when higher spin symmetry is slightly broken \refs{\MaldacenaSF,\GaberdielUCA}. 

\ifig\crossing{We consider the operator product expansion of the four-point correlator in the $u$- (left) or $v$-channel (right). The result should not depend on the order in which we are doing the OPE. This is the content of the crossing equation.} 
{\epsfxsize4.0in\epsfbox{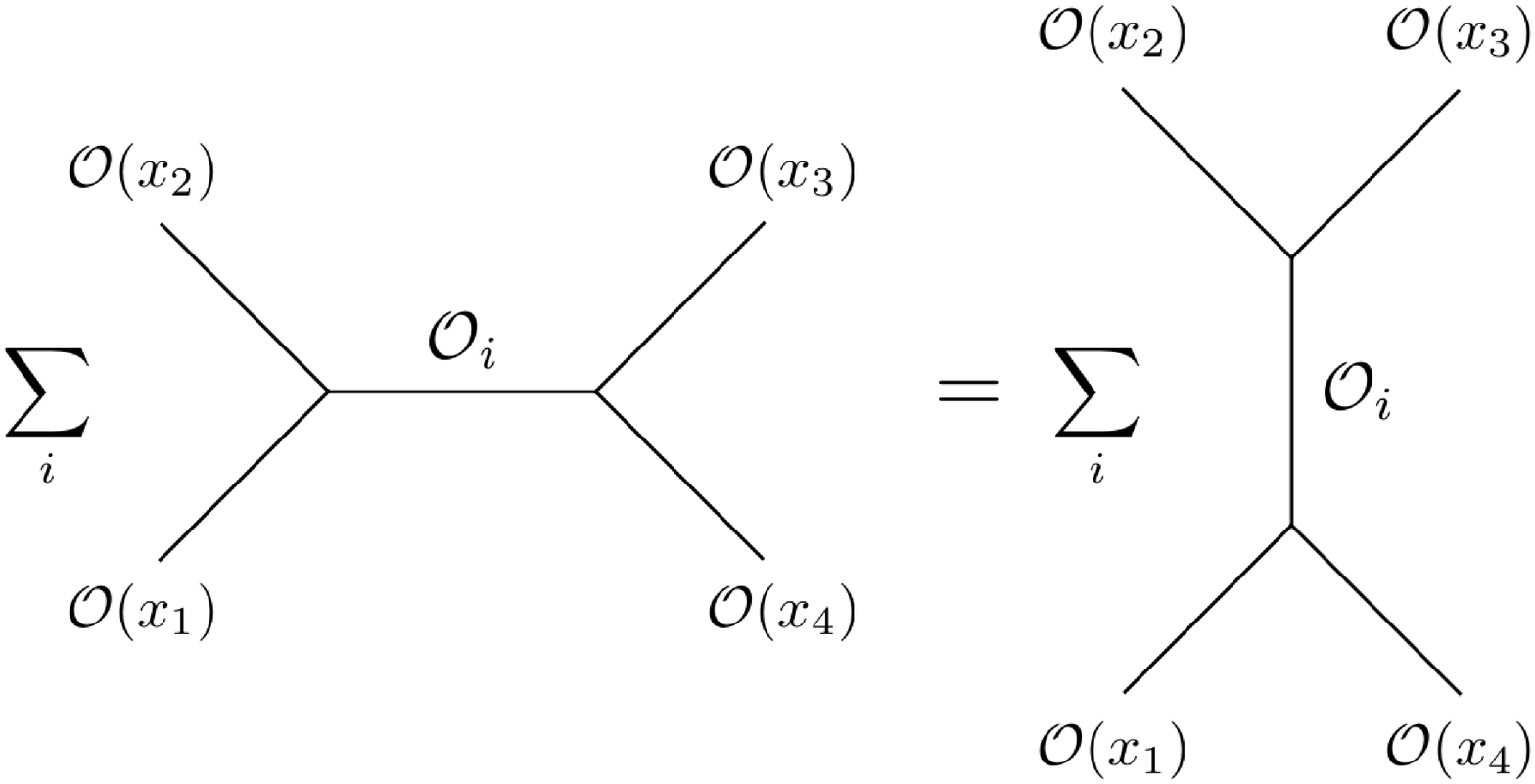}}

Conformal bootstrap is the simple idea that by imposing associativity of the operator product expansion (OPE) for local operators in a conformal field theory (see \crossing) and unitarity one, at least in principle, should be able to find the spectrum of local operators and their three-point couplings \refs{\FerraraYT, \PolyakovHA }. This knowledge is then enough to find all the correlation functions of local operators in the theory.\foot{An immediate hint that it should not be too easy is that through AdS/CFT that would mean solving, at least partially, the problem of quantum gravity in asymptotically AdS spaces and string theory in particular \refs{\MaldacenaRE\GubserBC-\WittenQJ} . A more humble but still unreachable goal would be a precise determination of critical exponents for the second order phase transitions in three dimensions \PolyakovXD.} 

Our strategy will be to focus on the double light-cone limit $u,v \to 0$ of the four-point correlation function of scalar operators.\foot{We introduce our notations at the beginning of the next section.} When ${u \over v}$ is kept fixed it is not known what is the behavior of the correlator in a generic CFT in this limit.\foot{In the context of gauge theories this limit was analyzed in \AldayZY .} 

First, we notice that the double light-cone limit is very simple in free theories. Four-point correlation functions admit an OPE-like expansion simultaneously in $u$ and $v$ with powers being controlled by the twists of the operators that are being exchanged in the $u-$ and $v-$ channel correspondingly. The crossing equation is then straightforward to solve and it splits into many pieces which satisfy crossing separately. These pieces correspond to an infinite number of operators of twist $\tau_1$ and $\tau_2$ which are dual to each other under crossing. 

Second, we observe that exactly the same structure is inherited by theories where higher spin symmetry is slightly broken. More precisely, we observe the same type of expansion as the one in the free theory, but dressed with polynomials of $\log u$ and $\log v$. The degree of the polynomial depending on the order of perturbation theory. 

Thirdly, by focusing on the lower twist operators in the double light-cone limit expansion we can fix the large spin limit behavior of the anomalous dimension of the higher spin currents $\gamma_s$. 

In section 2 we review the basic idea of our approach and discuss the double light-cone limit in free theories. In section 3 we analyze the perturbed crossing equation to leading order in the perturbation parameter. In section 4 we generalize the results of section three to all orders in perturbation theory. In section 5 we apply our physical picture to the 3d Ising model and predict the scaling dimensions of higher spin currents. In section 6 we present some conclusions and open problems. We also include several appendices with technical results and further examples.

\newsec{Basic Idea}

In this section we explain our approach to the crossing equation. The basic idea is similar in spirit to the works \refs{\FitzpatrickYX\KomargodskiEK-\AldayCWA} but generalizes them in several ways. 

\ifig\crossratios{By doing a conformal transformation we can always bring four points to the Minkowski plane. We depict the patch of two-dimensions Minkowski space. The cross ratios $z$ and $\bar z$ have an interpretation of the position of the second point $x_2  = (z,\bar{z})$. The other three points are chosen to be at $x_1  = (0,0)$, $x_3  = (1,1)$, $x_4  = (\infty,\infty)$.} {\epsfxsize4.0in\epsfbox{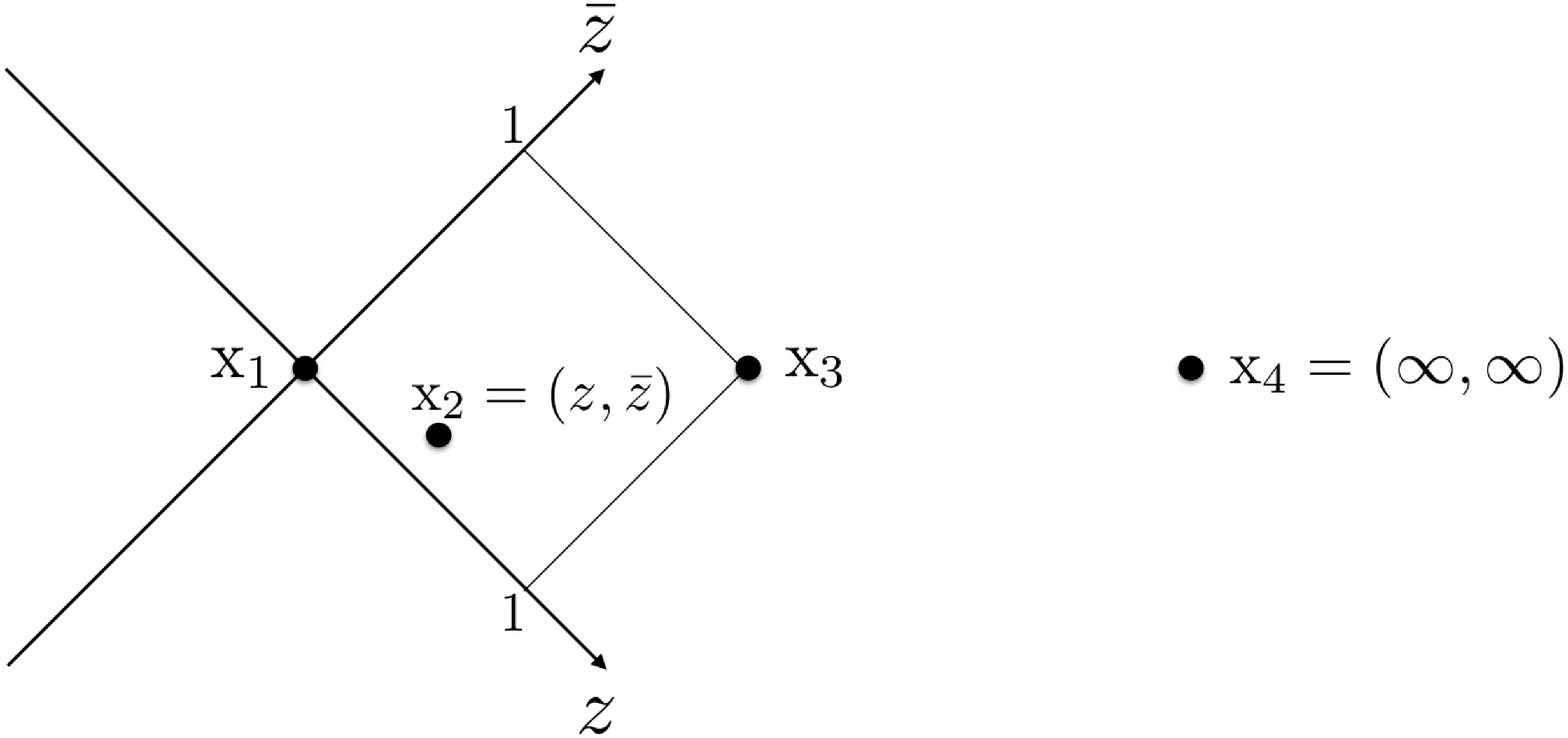}}

Our main object of interest is the four-point function of scalar operators in a CFT
\eqn\correlator{\eqalign{
\la {\cal O} {\cal O} {\cal O} {\cal O} \ra &=  {G(u,v) \over (x_{12}^2 x_{34}^2)^{\Delta} } , \cr
u= z \bar z = {x_{12}^2 x_{34}^2 \over x_{13}^2 x_{24}^2} &,~~~ v=(1-z)(1- \bar z) = {x_{14}^2 x_{23}^2 \over x_{13}^2 x_{24}^2},
}}
where as usual $z$ and $\bar z$ can be interpreted as light-coordinates in the Minkowski plane \crossratios.

The crossing equation takes the form
\eqn\crossing{
f(u,v) = v^{\Delta} G(u,v) = u^{\Delta} G(v,u)=f(v,u).
}
The nontrivial content of \crossing\ is that each side can be decomposed in terms of conformal blocks with positive coefficients in front. This is implied by the fact that the theory is unitary and admits an OPE expansion.

We will be interested in the double light-cone limit $u,v \to 0$ or $z \to 0$, $\bar z \to 1$. On the picture above \crossratios\ this limit corresponds to sending the point $2$ to the upper corner of the space-like separation diamond. There are three basic ways to take the limit. First, by taking the $u \ll v \ll 1$ limit we are in the regime of the $u$-channel light-cone OPE which is dominated by the exchange of the lowest twist operators propagating in the $u$-channel. Second, by taking the $v \ll u \ll 1$ limit we are in the regime of the $v$-channel light-cone OPE which is controlled by the operators of the lowest twist propagating in the $v$-channel. Thirdly, by taking the limit such that $u \sim v \ll 1$ we are in the double light-cone regime. A priori it is not controlled by any OPE. Although this limit was discussed in \AldayZY\ in the context of gauge theories, the behavior of correlators for a generic CFT in this limit is, to the best of our knowledge, completely unexplored.  
 
The first observation that underlies our paper is that the double light-cone limit is very simple in free theories. It smoothly interpolates between the $u$-channel light-cone OPE and the $v$-channel light-cone OPE, with each of them being valid for any ratio ${u \over v}$. This makes the solution of the crossing relation \crossing\ straightforward since we can make the light-cone expansion in both channels simultaneously such that the correlator takes the form
\eqn\doubleOPE{
f(u,v) = \sum_{i,j} c_{i j} u^{{ \tau_i \over 2} } v^{{ \tau_j \over 2} } \ ,
}
where the sum goes over the twists of operators that are present in the free theory.\foot{The same holds for generalized free fields.} Crossing symmetry simply implies 
\eqn\doubleOPEb{
c_{i j} = c_{j i} \ .
}
We discuss the physics of this expansion below. 

The second observation is that when we depart away from the higher spin symmetric point the same picture holds. Namely if we work in the theory with some abstract coupling $g$ then the correlation function at order $L$ takes the form \doubleOPE\ with $c_{i j} = c_{i j}(\log u, \log v)$ being polynomials of order $L$ in $\log u$ and $\log v$. Crossing then is the statement that
\eqn\doubleOPEc{
c_{i j} (\log u, \log v)= c_{j i} (\log v, \log u) \ .
}
We claim that this is the most generic form of the correction to the correlation function at any order of perturbation theory in any weakly coupled CFT.

The third observation is that each term in the expansion \doubleOPE\ has a clear microscopic origin both in the $u$- and the $v$-channel. Thus, we can identify particular subsets of operators that obey crossing in the double light-cone limit. As we will explain below this picture of the double light-cone expansion of correlators has very interesting implications for the spectra of higher spin currents.

\subsec{Review Of The Light-Cone OPE Regime $u \ll v$}

Let us first review the regime $u \ll v \ll 1$. In this regime the correlation function above is controlled by the light-cone OPE in the $u$-channel so that we can write
\eqn\OPEuchannel{
G(u,v) \approx 1 + u^{{\tau_{{\rm min}} \over 2} } F_{\tau_{{\rm min}}}(v) + ...
}
where in the RHS we kept only the contribution of the leading twist operators. 
The basic observation of \refs{\FitzpatrickYX,\KomargodskiEK} was that due to \crossing\ and the convergence of the $v$-channel OPE the function of $v$ that appears in\OPEuchannel\ should come from twists of physical operators exchanged in the $v$-channel
\eqn\basicfact{
v^{\Delta} \left( 1 + u^{{ \tau_{{\rm min}} \over 2} } F_{\tau_{{\rm min}}}(v) + ... \right) =u^{\Delta} \sum_{\tau_i} v^{{ \tau_i \over 2} } F_{\tau_{i}} (u) .
}

Requiring \basicfact\ leads to certain constraints on the spectrum. In particular, \refs{\FitzpatrickYX,\KomargodskiEK} considered a generic CFT where the leading twist operator is separated from all the other operators by a twist gap. This typically happens in strongly coupled CFTs. It was then shown that the presence of the unit operator leads to the prediction of an infinite set of double trace-like operators with their twist approaching $2 \Delta + n$ at infinite spin $s$, whereas the correction due to the exchange of the minimal twist operator in the $u$-channel is captured by the anomalous dimension $\gamma (n, s)$ of the double twist operators and their three-point functions, see also \AldayMF, \KavirajCXA.

One natural question to ask is if this mechanism is applicable to weakly coupled CFTs as well. The immediate difference is that in weakly coupled CFTs we have an infinite tower of almost conserved higher spin currents which are not separated from the stress tensor by a twist gap. As we will show this leads to a qualitative change of the physical picture above. Understanding the way the crossing equation is satisfied in this case is the main subject of this paper. 

Another obvious problem with applying the results of \refs{\FitzpatrickYX, \KomargodskiEK}\  to 
weakly coupled theories is that they are expected to hold for spins $s$ that satisfy
\eqn\condition{
\delta \tau_{gap} \log s \gg 1 ,
}
where $\delta \tau_{gap}$ stands for the twist gap between the stress tensor and the next operator. In weakly coupled theories $\delta \tau_{gap} \ll 1$ and thus the condition \condition\ is satisfied only for operators with exponentially large spins which is not what one can compute using perturbation theory or numerical methods. For example, in the 3d Ising model $\delta \tau_{gap} \sim 0.02$. The require spin to make $\delta \tau_{gap} \log s \gg 1$ would be quite far from the recent numerical bootstrap results, where  anomalous dimensions of higher spin currents were extracted up to $s \simeq 40$ \ElShowkDWA.

For weakly coupled CFTs it is more natural to consider exactly the opposite regime 
\eqn\conditionB{
\delta \tau_{gap} \log s \ll 1 ,
}
This regime was analyzed in \AldayCWA ~in the context of four-dimensional maximally super-symmetric Yang-Mills. There it was noted that in perturbation theory the contribution from leading-twist operators of high spin to the four-point correlator is of the form
\eqn\Gab{
G(u,v) \sim {u \over v} h(\log u,\log v)
}
to any order in perturbation theory and where $h(\log u,\log v)$ is a symmetric function as a consequence of crossing.
This was then used to derive the behavior at large spin  of the OPE coefficients of higher spin currents.

Understanding the regime \conditionB\ for general CFTs, with the focus on anomalous dimensions of higher spin currents $\gamma_s$, is the subject of the present paper. This generalizes the analysis of \AldayCWA, as well as unifies \refs{\FitzpatrickYX\KomargodskiEK-\AldayCWA} into a single framework. Understanding this regime is not only relevant for theories describing critical phenomena in three dimensions but also it allows comparison to predictions of the bootstrap methods with the results available through perturbative analysis as well as numerical methods. 

\subsec{Double Light-Cone Expansion in Free Theories}

In this section we would like to consider how the crossing equation works in free theories in the double light-cone limit. 
In free theories this limit is very simple and is controlled by the exchange of free particles along both light cones. It is also important to emphasize that since we approach the light-cone from the space like direction the OPE is convergent in both channels.

To take the limit we can imagine rescaling $(u, v) \to \lambda (u,v)$ in \crossing\ and expanding the correlation function in $\lambda$. In free theories the result for any scalar correlator is very simple: we get a polynomial in $\lambda$ of degree fixed by $\Delta$. Recall that in the free theory $\Delta$ can either be integer or half-integer.\foot{Here we assume that the dimensionality of spacetime is integer but analogous discussion holds in any $d$.} In this way we get

\eqn\resultfree{\eqalign{
f(\lambda u, \lambda v) &=\sum_{i} \lambda^{i \over 2} f_{i} (u,v), \cr
f_{i}(u,v) &= f_i (v,u) ,
}}
where the second equation is just the statement of crossing and $f_{i} (u,v)$ are homogeneous symmetric polynomials in $\sqrt{u}$ and $\sqrt{v}$ of degree $i$. A convenient basis for such polynomials is given by 
\eqn\basis{
g^{i}_j = u^{{ j \over 2} } v^{{i - j \over 2} } +  v^{{ j \over 2}} u^{{i - j \over 2} }  .
}

In other words, the correlator in free theory admits an expansion in twists in both direct and dual channels
\eqn\doubletwist{
f( u, v) = \sum_{m,n} c_{m n} u^{{m \over 2}} v^{{n \over 2}}, ~~~ c_{mn} = c_{nm}.
}

By expanding \basis\ at small $u$ we see that it consists of operators with twists $j$ and $i - j$ which are transformed into each other under the crossing transformation. Applying to this polynomial the logic of \basicfact\ we see that in the double light-cone limit the crossing equation for free theories is satisfied separately by pairs of twist trajectories\foot{As we explain below for operators of low twist the twist trajectory can be made out of one, or a finite number,  of conformal primary Regge trajectories. For high enough twists though the number of Regge trajectories that can contribute depends on spin.} which are mapped into one another by crossing. Symbolically, we can write it as follows
\eqn\crossingR{
{\rm twist}_1 \leftrightarrow {\rm twist}_2 .
}

This is different from the picture that emerged in \FitzpatrickYX, \KomargodskiEK\ where the stress tensor itself was mapped to the Regge trajectory of the double trace-like operators. Another and more dramatic difference which we explicitly demonstrate below is that generically higher spin currents are not mapped to the double trace-like operators in free theories at all.

To illustrate this idea let us consider for example the operator $\phi^2$ in the free scalar theory in $d$ dimensions. In this case we have \DolanUT
\eqn\funccrossrat{
G^{(0)}(u,v) = {u^{d-2}+v^{d-2}+u^{d-2} v^{d-2} \over v^{d-2} } +{1 \over c} {u^{(d-2)/2} v^{(d-2)/2} + u^{d-2} v^{(d-2)/2}  + u^{(d-2)/2} v^{d-2}  \over v^{d-2}} .
}
where $c$ is proportional to the central charge or two-point function of stress tensors in the theory. This has an expansion as in \resultfree\ with
\eqn\reggethings{\eqalign{
f_0 (u,v) &= u^{d-2} + v^{d-2} + {1 \over c} u^{{d-2 \over 2}} v^{{d-2 \over 2}}, \cr
f_{d-2} (u,v) &= {1 \over c} \left( u^{d-2} v^{{d-2 \over 2}} + u^{{d-2 \over 2} } v^{d-2} \right), \cr
f_{2(d-2)} (u,v) &= {1 \over c} u^{d-2} v^{d-2}.
}}

It is instructive to understand the microscopic interpretation of these polynomials. Recall that in the light-cone OPE, say $u \to 0$, operators with twist $\tau$ enter as $u^{{\tau \over 2}}$. 
In this way we can interpret $u^{d-2} + v^{d-2}$ as the unit operator with twist $\tau = 0$ and double trace-like operators with twist $\tau =2 \Delta = 2 (d-2)$, which map to each other under crossing
\eqn\crossingRb{
{\rm Regge}_{{\rm unit} } \leftrightarrow {\rm Regge}_{{\rm double \ trace-like}} .
}
This particular result is completely general and holds for any CFT with $d>2$ as briefly reviewed in the previous section.

The piece ${1 \over c} u^{{d-2 \over 2}} v^{{d-2 \over 2}}$ is quite special since it is self-dual and moreover it consists of operators with twist $\tau = d-2$ which are higher spin currents. Thus, we conclude that in the $\phi^2$ correlator 
\eqn\crossingRc{
{\rm Regge}_{{\rm HS} } \leftrightarrow {\rm Regge}_{{\rm HS}} .
}

This is the mechanism that was at play in \AldayCWA. We will see that this self-duality is not as generic as \crossingRb . Indeed, had we considered the correlator of $\phi$'s, the higher spin currents would have been double trace-like operators, mapped through crossing to the unit operator. 

\subsec{Microscopic Realization}

Let us next understand the picture above microscopically. If we consider the expansion of $G(u,v)$ it takes the well-known form
\eqn\expansion{
G(u,v) = \sum_{\tau,s} a_s u^{{\tau \over 2}} g_{\tau,s} (v),
}
where we expanded the contribution of a given twist in terms of collinear conformal blocks. $\tau$ is the twist of the operators that appear in the OPE of ${\cal O} {\cal O}$, $s$ is their spin and $a_s$ are the squares of the corresponding three-point couplings. Focusing on the higher spin currents in the example above we get
\eqn\leadingtwist{\eqalign{
&\sum_{s=0}^{\infty} a_{s}^{(0)} \ _2 F_{1} ({d-2 \over 2}+ s, {d-2 \over 2}+s, d-2 + 2 s , 1-v)=  {1 \over c} {1 \over v^{ {d-2 \over 2} }}\left( 1 + O(v) \right), \cr
&a_{s}^{(0)} = {(1 + (-1)^{s} ) \over 2} {2 \over c} { \Gamma (s+{d \over 2} - 1)^2  \Gamma (s+ d -3) \over \Gamma(d/2 - 1)^2 \Gamma(s+1) \Gamma(2 s+d -3) } ,
}}
The crucial point is the following: recall that each conformal block has only a logarithmic singularity as $v \to 0$. Hence, if we are to reproduce the power-like singularity with the sum \leadingtwist\ we have to sum over an infinite number of operators. Moreover, such singularity must come from the ``tail'' of the sum, and must be insensitive to the change of the lower limit of the sum. In other words,  $\sum_{s=0}^{\infty}$ and $\sum_{s= s_0}^{\infty}$ should produce the same singularity. In our paper we will always focus on power-like singularities coming from the sum over an infinite number of spins.

To see how \leadingtwist\ is satisfied it is instructive to notice that as $v$ becomes small, the sum is dominated by operators of large spin $\sim {1 \over \sqrt{v}}$. As shown in \refs{\FitzpatrickYX, \KomargodskiEK}, the leading divergence of the l.h.s. of \leadingtwist\ can be computed by considering the scaling limit $s = {h \over \sqrt{v}}$ and converting the sum over spins into an integral over $h$
\eqn\resultA{\eqalign{
\sum_s \to {1 \over 2} \int dh \ ,
}}
where the ${1 \over 2}$ factor appears because only even spins contribute. 
In this limit the ingredients of \leadingtwist\ become \foot{This result can also be obtained by taking the scaling limit of the Casimir equation satisfied by the collinear conformal blocks \AldayZY. The method presented here, however, is more systematic. }
\eqn\resultA{\eqalign{
_2 F_1 ({h \over \sqrt{v}}, {h \over \sqrt{v}}, 2 {h \over \sqrt{v}}, 1 - v ) &\to \sqrt{ {h \over \sqrt{v} \pi} } 2^{ {2 h \over \sqrt{v}}} K_0 (2 h), \cr
a^{(0)}_{{h \over \sqrt v}} &\to {1 \over c} {\sqrt \pi \over \Gamma(d/2 - 1)^2}2^{2 - {2 h \over \sqrt v} }\left({h \over \sqrt v} \right)^{d - {7 \over 2}}
}}
from which the leading divergent piece can be computed 

\eqn\identity{
{1 \over c} {1 \over v^{d-2 \over 2}} {4 \over \Gamma(d/2 - 1)^2} \int_0^{\infty} d h \ h^{d-3} K_0 (2 h) = {1 \over c} {1 \over v^{d-2 \over 2}}.
}
which agrees precisely with the leading divergence of the r.h.s. of \leadingtwist\ . The lesson is that the dominant contribution comes from the operators with large spin. Changing the cross ratio makes different operators dominant. A similar exercise can be repeated for other cases. By re-summing operators of given twist and large spin we recover the dual twist operators in the other channel. The dominant spin in the $12$ channel is ${1 \over \sqrt{v}}$ whereas in the $23$ channel the dominant spin is ${1 \over \sqrt{u}}$. We demonstrate how similar expansion works in the 2d Ising and Yang-Lee models in appendix B.

Free theories are the only known examples of unitary CFTs with finite central charge and higher spin symmetry. If we allow for infinite central charge then there is another famous example which is the critical $O(N)$-model. In this situation the same logic applies. 

\newsec{Slightly Broken Higher Spin Symmetry}

Next we would like to depart from the free theory limit. When the higher spin symmetry gets broken conserved higher spin currents receive anomalous dimension $\gamma_s$ which we assume to be very small

\eqn\smallanom{
\tau_{s} = d-2+ \gamma_s,~~~ \gamma_s \ll 1.
}

The typical example would be to consider a Lagrangian theory and turn on an exactly marginal deformation. Examples of this type are SCFTs. We could also consider Wilson-Fischer $\phi^4$-model in $4 - \eps$ or $\phi^3$-model in $6 - \eps$ dimensions, where $\eps$ is a small parameter. Yet another example is the large $N$ limit of the Wilson-Fischer fixed point in $2< d < 4$, in which the role of small parameter is played by $1/N$. We can think about all of these parameters as the coupling $g$ and assume that $\gamma_s \sim g f(s)$ and similarly for the three-point couplings. 

Let us understand the new effects that we get as we analyze the crossing equation to first order in $g$. There are three of them:

- external operators receive correction to their dimension; 

- operators that were present in the OPE at tree-level receive corrections to their scaling dimensions, while the corresponding OPE coefficients get corrected as well;

- new operators that were absent at tree-level appear in the OPE.

Let us discuss how different effects manifest themselves at the level of the crossing equation.

\subsec{Case Of Protected Operators}

Let us  start by making two additional assumptions, both of which will be relaxed later: 

- external operators do not receive correction to their dimension; 

- there is no operator $\phi$ in the original spectrum. In other words, $\phi^2$ is the leading twist scalar operator. 

First, let us recall that the new operators that appear at leading order in $g$ enter with their tree-level dimension which is integer or half-integer. Second, the effect of operators present at tree-level in the OPE and receiving anomalous dimensions is very distinct. Indeed, we have
\eqn\anomdim{
u^{{\tau_0 + \gamma_{\tau_0, s} \over 2}} \approx u^{{\tau_0 \over 2}} \left(1 + { \gamma_{\tau_0, s} \over 2} \log u \right)
}
where we assumed $\gamma_{\tau_0, s} \log u \ll 1$ which, as will become clear below, is equivalent to \conditionB . Let us consider \basicfact\ to first order in perturbation theory and let us focus on the contribution coming from the correction to the anomalous dimension of the leading twist operators, which for the present case, are higher spin currents 
\eqn\basicfactpert{
v^{d-2} u^{{d-2 \over 2}} \log u \sum_{s} {\gamma_s \over 2} a_{s}^{(0)} \ f_{s} (v) = u^{d-2} \left(\sum_{\tau_i^{(0)}} v^{{ \tau_i^{(0)} \over 2} } \delta F_{\tau_{i}}^{(0)} (u) + \log v \sum_{\tau_i^{(0)},s} v^{{ \tau_i^{(0)} \over 2} } { \gamma_{\tau_i^{(0)},s} \over 2}  F_{\tau_{i}}^{(0)} (u)  \right),
}
where $a_s^{(0)}$ are the square of the OPE coefficients of the higher spin currents at $g=0$, as in the computation above. By $f_{s} (v)$ we denoted the collinear conformal block which again appeared in the computation above. Lastly, $\gamma_s$ stand for anomalous dimensions of higher spin currents. The first term on the RHS corresponds to the contribution from new operators, that were absent in the tree-level OPE, as well as corrections to the OPE coefficients of operators that were present at tree-level. Both contributions enter with the twists of the free theory $\tau^{(0)}_{i}$. The second term in the RHS is due to the anomalous dimensions of operators that were present in the original OPE.

Let us understand possible behaviors of the RHS for small $v$. We can renormalize our operators such that the correction to the unit operator is absent. Next, we can have the correction from the HS currents that corresponds to $\tau^{(0)}_{i} = d-2$. This leads to the following relation
\eqn\relation{
\sum_{s} {\gamma_s \over 2} a_{s}^{(0)} f_{s} (v) = {c_1 \log v + c_2 \over v^{{d-2 \over 2}}}\left(1 + O(v) \right),
}
where $c_{1,2}$ are constants which are fixed by the microscopical theory. A priori we do not know if they are zero or not. For now let us assume that they are nonzero. We will come back to the other case below. As discussed for the free theory, the divergence for small $v$ can only be reproduce by  the contribution from the higher spin operators. 
Turning the sum into an integral as before we get
\eqn\integral{
{1 \over 2} {4 \over c \ \Gamma(d/2 - 1)^2} \int_0^{\infty} d h \ h^{d-3} K_0 (2 h) \gamma({h \over \sqrt{v}})= c_1 \log v + c_2 .
}
In perturbation theory we can assume $ \gamma({h \over \sqrt{v}})$ to be a polynomial. In which case this has a unique solution
\eqn\result{\eqalign{ 
\gamma_s &= a_1 \log s + a_2, \cr
a_1 &= - 4 c \ c_1,~~~ a_2 = 2 c \left( c_2 + 2 c_1 \psi^{(0)}({d - 2 \over 2}) \right).
}}

In this simple example we see how the relation \crossingRc\ works in the perturbed theory. In the computation above we have assumed that the sum is dominated by large spins of order $1/\sqrt{v}$. We see that this is indeed the case, as the integral \integral\ is dominated by $h=h_* \sim 1$.

We can also consider the contribution from operators with higher tree-level twist $\tau^{(0)} = d-2 +\alpha$, $\alpha>0$. Provided $\alpha < d - 2$ they will contribute a divergent term, and the divergences can be evaluated as explained above. Their contribution will result in a term of the form
\eqn\relation{
\gamma_s = {c_{1, \alpha} \log s + c_{2, \alpha} \over s^{\alpha}} ,
}
which is sub-leading at large $s$. To summarize, assuming that there are no operators in the original theory inside the twist gap ${d-2 \over 2} \leq \tau < d-2$ we predict that the most general leading form of the anomalous dimension of higher spin currents takes the form
\eqn\genform{
\gamma_s = c_1 \log s + c_2 + O({1 \over s }) .
}

Below we will show that if $c_1 \neq 0$ then $\log s$ is the leading asymptotic to any order in perturbation theory. Of course, the typical examples of a theory like above are weakly coupled gauge theories. In all known cases \genform\ indeed is the correct form of the leading spin anomalous dimension of higher spin currents. As explained in \AldayMF\ it can be understood as coming from the color flux created by the gauge fields with $c_1$ being its energy density in appropriate coordinates. Above we demonstrated how this result emerges from the analysis of the bootstrap equation.

\subsec{Operators With Twist ${d - 2 \over 2} < \tau < d-2$}

In the consideration above we assumed that the original theory does not contain operators 
with twist $\tau < d - 2$.\foot{In higher dimensions we should be careful with our definition of twist but since we are dealing with the scalar operators and their OPE we care only about symmetric traceless tensors.} If we consider an arbitrary combination of free fields the only operator with twist $\tau < d-2$ that can appear in the OPE of two scalar operators is the free scalar field $\phi$ itself. 

Assuming that the original theory contains a free scalar field $\phi$ we can consider its four point function\foot{Here it is only important for us that it contains {\it at least} one free scalar field.}
\eqn\fourphi{
G^{(0)}(u,v) = {u^{{d-2 \over 2}} + v^{{d-2 \over 2}} + u^{{d-2 \over 2}} v^{{d-2 \over 2}} \over v^{{d-2 \over 2}}}.
}

In this case higher spin currents are dual to the unit operator and to themselves. The tree-level three-point functions are exactly the same as above if we set $c=1$. We consider 
a correction to this correlator that breaks higher spin symmetry. In this case \basicfactpert\ takes the form
\eqn\basicfactpertC{
v^{{d-2 \over 2}} u^{{d-2 \over 2}} \log u \sum_{s} {\gamma_s \over 2} a_{s}^{(0)} \ f_{s} (v) = u^{{d-2\over 2}} \left(\sum_{\tau_i^{(0)}} v^{{ \tau_i^{(0)} \over 2} } \delta F_{\tau_{i}}^{(0)} (u) + \log v \sum_{\tau_i^{(0)},s} v^{{ \tau_i^{(0)} \over 2} } { \gamma_{\tau_i^{(0)},s} \over 2}  F_{\tau_{i}}^{(0)} (u)  \right),
}

The second term in the RHS can only arise from either the unit operator or higher spin currents themselves, since these are the only operators present at tree level. As we explained the unit operator does not lead to any interesting effects. 

A new feature in this case is that due to dynamics the operator $\phi$ itself can appear in the first term in the RHS. Notice that in the original theory $\phi$ is odd under the $Z_2$ parity $\phi \to - \phi$. Thus, for the operator $\phi$ to appear in the RHS the theory should break $Z_2$ invariance. 

Let us first consider the case when the breaking of $Z_2$ does happen. Notice that for twists ${d - 2 \over 2} < \tau < d-2$ the contribution in the RHS is fully given in terms of a single conformal block: that of the operator $\phi$. Focusing on the leading contribution at small $v$ we get 
\eqn\integralSIM{
{1 \over 2} {4 \over \Gamma(d/2 - 1)^2} \int_0^{\infty} d h \ h^{d-3} K_0 (2 h) \gamma({h \over \sqrt{v}})= - {\Gamma({d-2 \over 2} ) \over \Gamma({d-2 \over 4}) ^2}  c_{\phi \phi \phi}^2  v^{{d-2 \over 4}}  .
}
which is solved by
\eqn\solution{
\gamma_s = - {2 \Gamma({d-2 \over 2})^3 \over \Gamma({d-2 \over 4})^4}{c_{\phi \phi \phi}^2 \over s^{{d-2 \over 2}} } +\cdots.
}
where the dots represent sub-leading corrections, suppressed by powers of $s$.  

Next let us consider the case of a theory that preserves $Z_2$. In this case the $\phi$ operator does not appear in the OPE and higher spin currents are the leading twist operators in the dual channel as well. The sum rule now takes the form

\eqn\relationB{
\sum_{s} {\gamma_s \over 2} a_{s}^{(0)} f_{s} (v) = c_1 \log v + c_2  + O(v) ,
}
We note the RHS is not power divergent, so all spins will contribute equally, and in particular we cannot apply the integral method above. Indeed, if we were to apply the saddle point method blindly we would obtain

\eqn\integralSIMb{
{1 \over 2} {4 \over \Gamma(d/2 - 1)^2} \int_0^{\infty} d h \ h^{d-3} K_0 (2 h) \gamma({h \over \sqrt{v}})= \left( c_1 \log v + c_2 \right) v^{{d-2 \over 2}}.
}
and the would be solution $\gamma_s =\tilde c_1 {\log s \over s^{d-2}} + \tilde c_2 {1 \over s^{d-2}}$ would produce an integral that diverges for small $h$. As we will explain in the next section, we can overcome this difficulty by introducing a new idea. At this point we can only say that 
\eqn\constraintgamma{
\gamma_s < {1 \over s^{d-2 + \eps}}
}
where $\eps > 0$. A decay at a smaller rate would produce a divergent term, which is not observed. 

\subsec{Anomalous Dimension of External Operators}

Here we would like to relax the assumption that the external operators stay protected. It is easy to see that the same arguments follow through. Indeed, consider the corrected crossing equation
\eqn\correctedcross{
v^{\Delta_0 + \gamma_{ext}} G(u,v) = u^{\Delta_0 + \gamma_{ext}} G(v,u)
}
where we included a potential correction to the dimension of the external operator $\gamma_{ext}$.
Expanding to leading order in $\gamma_s,\gamma_{ext}$, using the crossing equation and focusing on the $\log u$ in front of higher spin currents we get

\eqn\basicfactpertB{\eqalign{
&v^{\Delta_0} u^{{d-2 \over 2}} \log u \sum_{s} {\gamma_s - 2 \gamma_{ext}\over 2} a_{s}^{(0)} \ f_{s} (v) \cr
&= u^{\Delta_0} \left(\sum_{\tau_i^{(0)}} v^{{ \tau_i^{(0)} \over 2} } \delta F_{\tau_{i}}^{(0)} (u) + \log v \sum_{\tau_i^{(0)},s} v^{{ \tau_i^{(0)} \over 2} } { \gamma_{\tau_i^{(0)},s} - \gamma_{ext} \over 2}  F_{\tau_{i}}^{(0)} (u)  \right),
}}
so the effect of the anomalous dimension for external operators in the discussions above is the shift $\gamma_s \to \gamma_s - 2 \gamma_{ext}$. 

If we are to consider higher order corrections we should also take into account that the saddle point that reproduces the unit operator shifts slightly \refs{\FitzpatrickYX\KomargodskiEK}. This is taken into account by correcting the three-point functions. This is used in one of the examples below.

\newsec{Generalization to All Orders in The Breaking Parameter}

The leading order analysis of the previous section suggests the following picture.
When we are in the regime where the $u$-channel OPE is valid we see that the tree-level expansion in terms of the twists of the free theory $u^{\tau^{(0)}/2}$ gets modified in two ways. 
First, we get new operators in the OPE that were absent at zeroth order. These will again enter with the tree-level twist. Second, the operators that were already present in the OPE get anomalous dimensions which lead to the emergence of terms of the type $u^{\tau^{(0)}/2} \log u$.

Of course, exactly the same things can be said about the $v$-channel. Assuming that the interpolation between the two channels is smooth as it was in the free theory we get the following generalization of \doubletwist\

\eqn\doubletwistC{\eqalign{
f( u, v) &= \sum_{m,n} c_{m n}(\log u, \log v) u^{{m \over 2}} v^{{n \over 2}}, ~~~ c_{mn}(\log u, \log v) = c_{nm} (\log v, \log u)\ ,\cr
c_{m n} &= c_{m n}^{(0)} + g \left(c_{m n | 0 0}^{(1)} + c_{m n | 1 0}^{(1)} \log u + c_{m n | 0 1}^{(1)}\log v + c_{m n | 1 1}^{(1)} \log u \log v \right) .
}}

We propose that to an arbitrary $L$-th order of perturbation theory the correlator takes the following form
\eqn\doubletwistD{\eqalign{
c_{m n}^{(L)} &=g^L \sum_{i,j=0}^{L }c_{m n | i j}^{(L)} ( \log u)^i (\log v)^j , \cr
c_{m n | i j} &= c_{n m| j i} .
}}

This ansatz is manifestly consistent with both $u$- and $v$-channel OPE and satisfies crossing. Also the structure of this ansatz agrees with the one expected from the evaluation of Feynman diagrams. Furthermore, it is shown in appendix A that perturbative results present in the literature admit the expansion \doubletwistD. See also appendix B for some simple 2d examples.

Of course, it is very easy to write an ansatz which is consistent with $u$- and $v$-channel OPE in 
an abstract CFT which takes the form
\eqn\ansatzgen{
f( u, v) = \sum_{m, n} c_{m n} u^{\tau_m/2} v^{\tau_n/2}, ~~~ c_{m n} = c_{n m},
}
where the sum goes over exact twists in the theory. It would be very interesting to understand what 
are the necessary conditions for the formula \ansatzgen\ to be valid in a generic CFT.\foot{It is well-known that for multi-variable functions real analyticity in each variable does not necessarily imply analyticity in both. As a trivial example consider $$f(u,v)= {u v \over u^2 + v^2}$$ which is crossing-symmetric and has OPE-like expansion both in the $u$- and the $v$-channel. However it cannot be written in the form \ansatzgen.} 

\subsec{On The Degeneracy Of Operators With High Twist}

It is true that in the expansion above \ansatzgen\ particular terms can be identified with the microscopic contribution of operators with given twist either in the $u$- or in $v$-channel . Thus to understand the utility of the expansion \ansatzgen\ we need to understand the structure of the spectrum of operators with a given twist. 

For us the relevant counting comes from considering the theory of a scalar field $\phi$. The fixed twist $\tau = n {d-2 \over 2}$ spectrum is controlled by fixing the number of fields $\phi$ and computing the number of ways we can distribute $s$ derivatives over them. In this way we can understand the number of primaries with the given twist $\tau$ as a function of spin $s$. More precisely, the number of primaries is given in terms of number of partition of $s$ into $n$ parts minus the number of descendants coming from the $(s-1)$-level
\eqn\primaries{\eqalign{
N_{n, s} &= p_{s}(n) - p_{s-1}(n) ,  \cr
\sum_{s=0}^{\infty} N_{n, s} x^s &= {1 \over x^2} \left( {1 - x \over (x,x)_{n}} - 1 \right),
}}
where $(x,x)_{n}$ is the $q$-Pochhammer symbol. 

An important point is that for large spin we have the following asymptotic behavior
\eqn\asympt{
N(n,s) \sim {s^{n-2} \over \Gamma(n-1)\Gamma(n+1)} \ ,
}
so that the number of primary operators does not grow in the case of higher spin currents, which correspond to $n=2$. On the other hand, for $n>2$ we see that the number of primary operators grows with the spin. This makes the utility of the expansion \ansatzgen\ much harder when we focus on higher twist operators. In this case we cannot focus only on finite number of Regge trajectories.\foot{In strongly coupled theories we can consider double trace-like operators of higher twists, of the form schematic form ${\cal O} \partial^s \square^n {\cal O}$. Their number does not grow with the spin. Many interesting properties for their anomalous dimension and OPE coefficients have been uncovered in \refs{\KavirajCXA,\KavirajXSA} .}
Our paper is bounded to the analysis of the low twist operators which lie on the finite number of Regge trajectories. It would be very interesting to understand what are the properties of the twist spectrum in a generic CFTs.

\subsec{All-Order $\log s$ Behavior}

In this section we would like to establish the leading $\log s$ behavior \genform\ to all orders in perturbation theory. A general argument for this behavior, based on symmetry arguments, was given in \AldayMF. Below we will show this also follows neatly from our arguments.

Recall that as in the example above we are dealing with the self-dual term in the double twist expansion $u^{{d-2 \over 2} } v^{{d-2 \over 2} } h(\log u, \log v)$. At finite loop order $h(\log u, \log v)$ is a polynomial in $\log u$ and $\log v$. Microscopically, it is generated by corrections to anomalous dimensions and three-point functions of the higher spin currents. More precisely, $\log u$ terms are generated by the expansion of $u^{\gamma_s/2}$ in the small anomalous dimension parameter $\gamma_s$, whereas $\log v$ terms can come both from sum over spins or corrections to the three-point coupling. The sum rule becomes
\eqn\sumovergener{\eqalign{
\sum_{s} a_s (g) u^{{\gamma_s(g) \over 2}} f_s (v) &= {h(\log u, \log v) \over v^{d-2 \over 2}} , \cr
h(\log u, \log v)  &= h(\log v, \log u) . 
} }
where this equation is understood order by order in perturbation theory. In the previous section we discussed the solution of this equation to first order in perturbation theory. Here we generalize this analysis to arbitrary order.

The structure of $h(\log u, \log v)$ implies the following large $s$ expansions 
\eqn\logsansatz{\eqalign{
\gamma_s &= \gamma^{(1)} \log s + \gamma^{(2)} \log^2 s + \gamma^{(3)} \log^3 s + ... , \cr
{a_s \over a_s^{(0)} } &= 1 + a^{(1)} \log s + a^{(2)} \log^2 s + a^{(3)} \log^3 s + ... \ ,
}}
plus power suppressed terms and where $a^{(i)}$ and $b^{(i)}$ are arbitrary functions of the coupling which start at order $g^i$ or higher.

Applying the saddle point method and focusing on the dominant contribution we get
\eqn\integralusual{
{4 \over \Gamma({d \over 2} - 1)^2}\int_0^{\infty} d h \ h^{d-3} u^{{1 \over 2} \gamma_{{h \over \sqrt{v}}}} \left( {a_{ {h \over \sqrt v} } \over a_{ {h \over \sqrt v} }^{(0)} } \right) K_{0} (2 h) = h(\log u, \log v) .
}

Plugging the expansion \logsansatz\ into \integralusual\ and requiring order by order $ h(\log u, \log v)$ be symmetric we find the following simple result
\eqn\resulsolution{\eqalign{
\gamma_s &= \gamma^{(1)}(g) \log s, \cr
{a_s \over a_s^{(0)} } &= {\Gamma({d \over 2} - 1 - {\gamma_s \over 2}) \over \Gamma({d \over 2} - 1)^2} ,
}}
which establishes the logarithmic behavior to all orders in the perturbation parameter. 
If we focus now on the effect of the corrections to the behavior of the tree-level result $u^{{d-2 \over 2}} v^{{d-2 \over 2}}$ we see that it gets dressed by the factor $e^{- f(g) \log u \log v}$ times a sub-leading contribution. This is the same conclusion as the one obtained in \AldayZY .

\subsec{Acting With A Casimir Operator}

Next we focus on a theory with $Z_2$ symmetry where the lowest twist scalar operator is $\phi$ itself. Let us explain how one can cope with the difficulty that we encountered above. The idea was explained in \AldayEYA\ and basically consists in acting with the Casimir operator on both sides of the crossing equation. The relevant part of the Casimir operator takes the following form \AldayEYA
\eqn\casimird{
{\cal D} = (1- v)^2 \pa_v - u (1-v) \pa_u + v (1 - v)^2 \pa_v^2 + v u^2 \pa_u^2 - 2 u v (1-v) \pa_u \pa_v \ .
}
This differs from the Casimir operator considered in \DolanDV\ by the piece that acts trivially on the collinear conformal block. 

Collinear conformal blocks are eigenfunctions of this operator
\eqn\collinear{\eqalign{
f_{coll}^{\tau , s} &= u^{{\tau \over 2}} (1-v)^s \ _2 F_1 ({\tau \over 2} + s, {\tau \over 2} + s , \tau + 2 s , 1-v), \cr
{\cal D} f_{coll}^{\tau , s} &= {1 \over 4}(2 s+\tau)(2s+\tau-2)  f_{coll}^{\tau , s} .
}}

For us it is important that the eigenvalue behaves as $s^2$ for large spin. When applying the Casimir to a $u$-channel expansion, the contribution from high spins will become much more important, and this makes the saddle approximation valid for cases for which it was not valid before.\foot{Since the OPE converges exponentially fast \PappadopuloJK , the  application of the Casimir does not spoil its convergence.}

To understand the effect of this let us consider again the case of four scalar operators $\la \phi \phi \phi \phi \ra$ in the $Z_2$-preserving case. The terms in the double twist expansion which are relevant for the question of anomalous dimensions of higher spin currents are
\eqn\relevantterms{
f(u,v) \sim u^{{d-2 \over 2}} v^{d-2 \over 2}\left[ c_{d-2,d-2}[\log u, \log v] +  c_{d-2,d-1}[\log u, \log v] \sqrt{v} + ... \right].
}

Acting on this with the Casimir and focusing on the terms that contain $\log u$ we notice that the
term $c_{d-2,d-2|1k}$ generates a ${1 \over v}$ singularity
\eqn\actioninthev{
{\cal D} G(u,v) =  {\cal D} \left[ v^{{2-d \over 2}} f(u,v) \right] \sim  {\cal D} \left[ u^{{d-2 \over 2} } \log u \ (\log v)^{k} \right] \approx {k (k-1) u^{{d-2 \over 2} } \log u \ (\log v)^{k-2}  \over v} ,
}
notably this contribution starts only from the second order $k=2$. Microscopically this contribution comes from the anomalous dimensions of currents $\gamma_s$. Repeating the same steps as before we get that to reproduce each of these terms we have to solve the following equation
\eqn\integralSIMb{
{1 \over 2} {4 \over \Gamma(d/2 - 1)^2} \int_0^{\infty} d h \ h^{d-3} \left( {h^2 \over v}\right) K_0 (2 h) \gamma({h \over \sqrt{v}})= (\log v)^{k-2} v^{{d-4 \over 2}},
}
where the extra factor in the integrand came from the Casimir eigenvalue at leading order. Note that now the integral converges for small values of $h$. This can be solved by $\gamma_s \sim {(\log s)^{k-2} \over s^{d-2}}$. Thus, we predict that the leading behavior of the anomalous dimensions of higher spin currents takes the following form
\eqn\dimens{\eqalign{
\gamma_s &= {\alpha_0 (g) + \alpha_1 (g) \log s +  \alpha_2 (g) \left( \log s \right)^2 + ... \over s^{d-2}} \ , \cr
\alpha_i (g) &\propto g^{2+i} \left[ 1 + O(g) \right] .
}}

The theories that fall into this class are $O(N)$ critical models in $2 < d < 4$ dimensions including the 3d Ising model as we discuss in detail below.

\subsec{All-Order ${1 \over s^{\Delta_{\phi}}}$ Behavior}

Next we consider the case of the four-point function of scalar operators $\phi$ in the case when interactions break $Z_2$-symmetry. Generically, we expect the operator $\phi$ to appear in the OPE and it will be the leading twist operator. In the double twist expansion at tree-level $\phi$ does not appear and we have only the standard terms of the type $u^{{d-2 \over 2}} v^{{d-2 \over 2}}$. When we turn on the interactions the operator $\phi$ appears in the OPE. The corresponding contribution to the four-point function takes the following form
\eqn\contribution{
u^{{d-2 \over 2}} v^{{d-2 \over 4}} h(\log u, \log v) + u^{{d-2 \over 4}} v^{{d-2 \over 2}} h(\log v, \log u) .
}
Interpreted from the perspective of one of the channels this should come from the operator $\phi$ Regge trajectory and higher spin currents. Due to conformal symmetry the contribution of the operator $\phi$ is completely fixed to be the one of its conformal block. Thus, we get the following equation
\eqn\contributionB{
u^{{d-2 \over 2}} v^{{d-2 \over 4}} h(\log u, \log v) = c_{\phi \phi \phi}^2 u^{\Delta_{\phi}} v^{{\Delta_{\phi} \over 2}} \ _2 F_{1} ({\Delta_{\phi} \over 2}, {\Delta_{\phi} \over 2}, \Delta_{\phi} , 1-u)\arrowvert_{small~u}}
from which we can find $h(\log u , \log v)$ to be
\eqn\resultforh{
h(\log u, \log v) = - {\Gamma({d-2 \over 2} + \gamma_{\phi}) \over \Gamma({{d-2 \over 2} + \gamma_{\phi} \over 2}) ^2}  c_{\phi \phi \phi}^2 u^{\gamma_{\phi}} v^{{\gamma_{\phi} \over 2}} \left( \log u + 2 \left[ \psi_0 ({{d-2 \over 2} + \gamma_{\phi} \over 2}) - \psi_0 (1) \right] \right) .
}

As before we should compare this function against the large spin integral
\eqn\integralusual{\eqalign{
{4 \over \Gamma({d \over 2} - 1)^2} &\int_0^{\infty} d h \ h^{d-3} u^{{1 \over 2}\gamma_{{h \over \sqrt{v}}}}  \left(  {a_{ {h \over \sqrt v} } \over a_{ {h \over \sqrt v} }^{(0)} } \right) K_{0} (2 h) = v^{- \gamma_{\phi}}+ v^{{d-2 \over 4} - \gamma_{\phi}} h[\log u, \log v] , \cr
}}
where in the RHS we also included the contribution due to exchange of the unit operator in the $v$-channel since it corrects the three-point functions. The equation \integralusual\ should be understood order by order in perturbation theory. Moreover, when going to higher orders we have to act with the Casimir operator on both sides of the equation to make the saddle approximation valid.

The solution of this equation takes the following form
\eqn\solutionphicube{\eqalign{
{a_{ s } \over a_{ s }^{(0)} } &= {\Gamma({d-2 \over 2} )^2 \over \Gamma({d-2 \over 2} + \gamma_{\phi})^2}s^{2 \gamma_{\phi}} \left( 1 +  \gamma_s  \left[ \psi_0 ({{d-2 \over 2} + \gamma_{\phi} \over 2}) - \psi_0 (1) \right] \right) , \cr
\gamma_{s} &= 2 \gamma_{\phi} -  {2 \Gamma({d-2 \over 2} + \gamma_{\phi})^3 \over \Gamma({{d-2 \over 2} + \gamma_{\phi} \over 2}) ^4}  {c_{\phi \phi \phi}^2 \over s^{{d- 2 \over 2} + \gamma_{\phi}}} .
}}

It coincides with the results of \refs{\FitzpatrickYX, \KomargodskiEK} . Indeed, even though the $\phi^3$-model is in the perturbative regime there is a large twist gap between $\phi$ and almost conserved higher spin currents. Thus, the non-perturbative results of \refs{\FitzpatrickYX, \KomargodskiEK}  should be readily applicable. Here we explicitly see how these results are consistent with the double twist expansion and appear order by order in perturbation theory.

\subsec{Constraints From Convexity and Large Spin Liberation}

One can ask if there are additional constraints on the anomalous dimensions of higher spin currents. A very general property of leading twist operators is convexity which was originally found in \NachtmannMR\ and recently reviewed in \KomargodskiEK . It states that the anomalous dimensions of the leading twist operators that appear in the OPE of Hermitian conjugate operators being plotted against their spin lie on a convex curve. For the examples considered above this property fixes signs of the leading term coefficients. In the case of $f(g) \log s$ behavior it implies that $f(g)>0$. For the $\gamma_s \sim {c \over s^{\alpha}}$ case it implies that $c<0$.

Another constraint comes from the fact that in the OPE of two operators 
of twists $\tau_1$ and $\tau_2$ there are operators which approach twist $\tau_1 + \tau_2$ at infinite spin \refs{\CallanPU,\FitzpatrickYX,\KomargodskiEK}. When these operators happen to be higher spin currents we can apply this condition! 

First, let us consider the case $\gamma_s \sim f(g) \log s$. Our analysis is only valid in the regime $f(g) \log s \ll 1$. Assuming that the $\log s$ behavior persists for higher spins we should conclude that double trace-like operators mentioned above are not higher spin currents. This is precisely what happens in gauge theories where at very large spin the leading twist operators are double trace operators. 

Second, in scalar theories which contain the $\phi$ operator higher spin currents are the double trace-like operators of \refs{\CallanPU,\FitzpatrickYX,\KomargodskiEK} and thus in this case we conclude that 
\eqn\asymptoticbehavior{
\lim_{s \to \infty} \gamma_s = 2 \gamma_\phi .
}

Thirdly, in \refs{\FitzpatrickYX,\KomargodskiEK} the leading correction to \asymptoticbehavior\ was found. The analysis of these papers is expected to be valid for spins such that the twist gap between the leading twist operator and the following one $\delta \tau \log s \gg 1$ which corresponds to exponentially large spins for weakly coupled theories. We can connect this behavior to the analysis above which is valid for $\delta \tau \log s \ll 1$. The interpolation should be smooth and respect convexity. 

\newsec{Application To The 3d Ising Model}

Here we would like to discuss what can we say about the critical $O(N)$ models and 3d Ising model as a limiting case of those. In our analysis they correspond to a model with a scalar operator $\phi$ and interactions which preserve the $Z_2$ symmetry. It is important to emphasize that we consider the correlator of four identical spin fields $\la S_1 S_1 S_1 S_1\ra$. The higher spin currents that appear in the tree-level OPE expansion of this correlator are the symmetric traceless currents and, thus, our analysis applies to the anomalous dimension of those.

First, let us review what is the prediction of the analysis of  \refs{\FitzpatrickYX,\KomargodskiEK}. Using their formulas for the stress tensor exchange we obtain
\eqn\predictionlarge{
\gamma_s - 2 \gamma_{\phi} \sim {\gamma_\phi^2 \over N \ s^{d-2}}, ~~~ \gamma_{\phi} \log s \gg 1 .
}
On the other hand, in the regime $ \gamma_{\phi} \log s \ll 1 $ the correction to the anomalous dimension of the currents due to higher spin currents in the dual channel should be of order $g^2$ or ${1 \over N^2}$, according to our analysis
\eqn\predictionlargeB{
\gamma_s - 2 \gamma_{\phi} \sim {1 \over N^2} {1 \over s^{d-2}}, ~~~ \gamma_{\phi} \log s \ll 1 .
}
This expectation is consistent with the results of~\LangGE,\LangZW\ who observe that the ${1 \over s^{d-2}}$ is indeed absent at the leading ${1 \over N}$ order. 

Our analysis of the bootstrap equation suggests that the leading correction to the anomalous dimensions of currents takes the form \dimens
\eqn\correctionBootstrap{
\gamma_s - 2 \gamma_{\phi} \sim {f(\log s) \over \ s^{d-2}} .
}
Moreover, \dimens\ states that $\log s$ corrections are suppressed by extra powers of the coupling. Since we expect that \predictionlargeB\ should turn into \predictionlarge\ as we increase the spin, we predict $\log s$ corrections at higher orders in perturbation theory. Notice also that there is an extra scale in the problem. Indeed, the symmetric traceless currents are known to receive the correction at the ${1 \over N}$-level
\eqn\predictionlargeC{
\gamma_s - 2 \gamma_{\phi} \sim {1 \over N} {1 \over s^{2}}, 
}
due to the exchange of the $\sigma$-field with $\Delta_{\sigma} = 2$ in the dual channel. In the region $1 \ll s^{4-d} \ll N$ the dominant contribution comes actually from the scalar field with the non-minimal twist and not from the higher spin currents.

Let us now turn to the 3d Ising model. In this case the $Z_2$-odd scalar operator of minimal twist is called $\sigma$ and the perturbation parameter $g$ is given by $g \sim \gamma_{\sigma} \sim 0.018$. To compare with the numerical bootstrap data we would like to consider spins of order $4 \leq s \leq 100$. In this regime it is not clear that the dominant contribution to the anomalous dimension of higher spin currents comes from the higher spin currents in the dual channel, of the form ${g^2 \over s}$. Recall that the Ising model contains a $Z_2$-even scalar operator in its spectrum, with $\Delta_{\varepsilon} \sim 1.41$. Since this scalar operator couples at order $g$ we expect this contribution to dominate, since ${g \over s^{\Delta_{ \varepsilon } }} \gg {g^2 \over s }$ for $s$ in the range of interest. Note that ${g \over s^{\Delta_{ \varepsilon } }} > {g^2 \over s }$ for roughly $s < 1.7 \cdot 10^4$.

The leading contribution of the scalar operator $\varepsilon$ is readily translated to the anomalous dimension of higher spin operators. The result 
is the following\foot{In order to derive this formula, one has to act with the Casimir operator, as explained above.} 
\eqn\toymodel{\eqalign{
\gamma_s &\simeq 2 \gamma_{\sigma}  - {2 \Gamma( \Delta_{ \varepsilon} ) \over  \Gamma( { \Delta_{ \varepsilon} \over 2} )^2 } {\Gamma (\Delta_{\sigma})^2 \over \Gamma(\Delta_{\sigma} - { \Delta_{ \varepsilon} \over 2} )^2} {f_{\sigma \sigma \varepsilon}^2 \over s^{\Delta_{\varepsilon} } } \ .
}}

We expect an infinite number of corrections to this result. First, we have a contribution from the higher spin currents of the type $- \gamma_{\sigma}^2 {c_0(\log s) \over s}$. Second, there are corrections due to the exchange of descendants of $\Delta_{\varepsilon}$, of the generic form ${\gamma_{\sigma} \over s^{\Delta_{\varepsilon} + n}}$. Lastly, there are corrections from the exchange of other heavy operators in the dual channel, the first one coming from the $Z_2$-even scalar $\varepsilon'$, with $\Delta_{\varepsilon'} \sim 3.83$.  We expect all these corrections to be very small for large values of the spin. As we explain bellow, $s=4$ is already in this regime! 

The corrections due to the descendants of the $\varepsilon$ scalar field (as well as to other heavier fields) can be evaluated following \AldayEYA.\foot{The correction to the anomalous dimension due to the $\Delta_{\varepsilon}$-exchange has the following form 
$$\delta \gamma_s =- {2 \Gamma( \Delta_{ \varepsilon} ) \over  \Gamma( { \Delta_{ \varepsilon} \over 2} )^2 } {\Gamma (\Delta_{\sigma})^2 \over \Gamma(\Delta_{\sigma} - { \Delta_{ \varepsilon} \over 2} )^2} { f_{\sigma \sigma \varepsilon}^2 \over j^{\Delta_{\varepsilon} }} \left(1
- \Delta_{\varepsilon}{ \Delta_{\varepsilon}^3 - 2 \Delta_{\varepsilon}^2 - (12 \Delta_{\sigma}^2-30 \Delta_{\sigma} + 10) \Delta_{\varepsilon} - 4  \over 24 (2 \Delta_{\varepsilon} - 1)} {1 \over j^2} +\cdots \right)$$
where $j^2 = (s - {1 \over 2} +{\gamma_{s} \over 2} )(s + {1 \over 2} +{\gamma_{s} \over 2} )$.}  
On the other hand, we cannot compute $c_0(\log s)$ from first principles, but, as argued above, its contribution should be small in the range of interest due to the $\gamma_{\sigma}^2$ factor. Using the the data of the numerical bootstrap $\Delta_{\sigma} = 0.518151$, $\Delta_{\varepsilon} = 1.41264$, $f_{\sigma \sigma \varepsilon}^2 = 1.10634$, \refs{\ElShowkDWA , \ElShowkHT, \SimmonsDuffinQMA} we arrive at the following formula
\eqn\toymodelB{\eqalign{
\gamma_s &\simeq 0.0363- {0.0926 \over s^{1.4126 }}+ {0.0012 \over s^{2.416}} - {0.0220 \over s^{3.4126}} - {c_0(\log s) \gamma_{\sigma}^2 \over s} \ .
}}

The first two terms in this expression are simply those of \toymodel . The next two come from the exchange of $\Delta_{\varepsilon}$. The last term comes from the higher spin currents.  We can estimate $c_0(\infty)$ by computing the contribution due to the stress tensor using the asymptotic formula of \refs{\FitzpatrickYX, \KomargodskiEK}. It gives $c_0(\infty) = 8.4988$ such that we get $- {0.0028 \over s}$ in the formula above. Moreover, from \dimens\ we expect that $c_0(\log s)$ is approximately constant $c_0$ for low spins. 

In this way we get the following anomalous dimensions for currents with low spin
\eqn\predictions{\eqalign{
\gamma_4^{c_0 = 0} &= 0.0231 , ~~~\gamma_6^{c_0 = 0} = 0.0289, \cr
\gamma_8^{c_0 = 0} &= 0.0314 ,~~~
\gamma_{10}^{c_0 = 0} = 0.0327 ,
}}
where we evaluated the anomalous dimensions using formula \toymodelB\ with $c_0 = 0$. For comparison let us also present the results for $c_0(\infty)$
\eqn\predictionsB{\eqalign{
\gamma_4^{c_0(\infty)} &= 0.0224 , ~~~\gamma_6^{c_0(\infty)} = 0.0284, \cr
\gamma_8^{c_0(\infty)} &= 0.0310 ,~~~
\gamma_{10}^{c_0(\infty)} = 0.0324.
}}

For constant $c_0$ we can also build combinations which are independent of it since the dependence on $s$ is simply inverse. In this way we get $c_0$-independent combinations
\eqn\predictiontwo{\eqalign{
\gamma_6 - {2 \over 3} \gamma_4 &= 0.0135, \cr
\gamma_8 - {1 \over 2} \gamma_4 &=0.0198, \cr
\gamma_{10} - {2 \over 5} \gamma_4 &= 0.0235.
}}

Our predictions are in agreement with the preliminary numerical bootstrap data of \ElShowkDWA . The result of \CampostriniAT\ for $\gamma_4 = 0.0208(12)$ is consistent with our prediction as well. We hope that further improvement of numerical methods can lead to a determination of anomalous dimensions of higher spin currents with higher precision. In particular precise determination of $\gamma_4$ may fix $c_0$ in the formulas above. 

We also expect \toymodel\ to work well for currents in the symmetric traceless representation of $O(N)$ symmetry in the $O(N)$ models with small $N$. It would be very interesting to compare this prediction with the numerical bootstrap results.

It would be also very interesting to check if $\log s$ terms appear at higher orders of perturbation theory. This would require computing higher order ${1 \over N}$ or $\eps$ corrections to the anomalous dimensions of the higher spin currents in the critical $O(N)$ model.

\newsec{Conclusions and Summary of Results}

In this paper we considered conformal field theories with almost conserved higher spin currents 
\eqn\higherspin{
\Delta_s = d-2+s + \gamma_s,~~~\gamma_s \ll 1.
} 

A typical example of those are weakly coupled conformal field theories, that is Lagrangian theories with exactly marginal deformations that admit a perturbative expansion. Another prominent examples are critical $O(N)$-models in various dimensions and the 3d Ising model. 

We analyzed the crossing equation in the double light-cone limit and solved for $\gamma_s$ at large spin $s$. In the double light-cone limit both cross ratios $u,v$ tend to zero with their ratio ${u \over v}$ being fixed. This limit is not known to be controlled by any OPE-like expansion in a generic CFT. In free theories we can analyze this limit explicitly and see it is an expansion in terms of the basic crossing-symmetric combinations
\eqn\basicblock{
u^{{\tau_1 \over 2}} v^{{\tau_2 \over 2}} + u^{{\tau_2 \over 2}} v^{{\tau_1 \over 2}} .
}
Microscopically, these come from an infinite set of operators with twist $\tau_1$ and $\tau_2$ which are mapped to each other by the crossing equation. 

When the higher spin symmetry is broken we can consider a perturbative expansion in $\gamma_s$ which is a small parameter in the problem. The expansion in twists \basicblock\ gets dressed by polynomials in $\log u$ and $\log v$ \doubletwistC, which roughly come from the expansion of $u^{\gamma}$ in the $u$-channel, or $v^{\gamma}$ in the $v$-channel. Moreover, the appearance of logarithms of higher degrees is what is expected from the evaluation of Feynman graphs. We also checked this expansion in the variety of examples (see appendices).

Focusing on the low twist operators and reproducing logarithmic corrections microscopically led to expressions for the anomalous dimensions of higher spin currents $\gamma_s$. The possible high spin behavior of those depends on the spectrum of the theory and symmetries of the problem.

We considered three basic cases. First, theories without scalar operators in the twist gap ${d-2 \over 2} \leq \tau < d-2$. In this case we found that the expected leading behavior at large spin is $\gamma_s = f(g) \log s$ \resulsolution. We showed that this result can be extended to all orders in the coupling. 
Second, we considered theories that do contain an almost free scalar operator $\phi$. We distinguish two classes of theories: those that break $Z_2$-symmetry $\phi \to - \phi$ and those that do not. For the $Z_2$-preserving case we argued that the expected expansion takes the form \dimens. A prominent example of a theory of this type are critical $O(N)$ models in various dimensions.  For the $Z_2$-breaking case the expected expansion takes the form \solutionphicube . Famous examples of this class are $\phi^3$ Yang-Lee type of models in various dimensions.

We considered also the application of our results to the 3d Ising model. In this case due to the additional suppression of the contribution of the higher spin currents that we explained in the case of the critical $O(N)$-model, what dominates the anomalous dimension for $s<10^4$ is the exchange of the sub-leading twist operator $\varepsilon$ in the dual channel. Using this fact we made a prediction for anomalous dimensions of higher spin currents $\gamma_s$ (see \toymodel, \toymodelB, \predictiontwo\ and the discussion thereby). We hope our prediction can be tested against the results of the numerical bootstrap in the near future.\foot{Our prediction is in agreement with the preliminary results of \ElShowkDWA.} Similar technique can be used to predict anomalous dimensions of certain higher spin currents in the critical $O(N)$ models with small $N$ and other scalar theories in various dimensions (see \refs{\FeiYJA, \ChesterGQA}).

A double twist expansion seems to be closely related to Mellin amplitudes with Mellin variables playing the role of complexified twists in the expansion above \PenedonesUE . Indeed, in the simple explicit examples one can go from the sum in the expansion to the contour integral as is familiar from the Regge theory. It would be nice to understand this relation better. It would also be interesting to explore the behavior of correlation functions in the double light-cone limit in two-dimensional CFTs more thoroughly.

We considered the scalar operators built out of scalar fields $\phi$ but similar analysis can be generalized to the scalar operators made of fermions or tensor fields. See appendix C.

Although the small $u,v$-expansion trivializes part of crossing, the full crossing symmetry cannot be imposed within the small $u,v$-expansion. One way to impose it is to use Mellin amplitudes. It would be nice to understand the implication of the full crossing symmetry for the small $u,v$-expansion. It can potentially lead to classification of all perturbative solutions of the crossing equation in the spirit of \refs{\HeemskerkPN, \AldayTSA}.

\newsec{Acknowledgments}

We would like to thank  B. Basso, A. Bissi, D. Cristofaro-Gardiner, T. Dumitrescu, Z. Komargodski, T. Lukowski, J. Maldacena, J. Penedones and S. Rychkov for useful discussions. We thank Slava Rychkov for sharing with us preliminary results of the numerical bootstrap evaluation of anomalous dimensions and three-point couplings of the higher spin currents in the 3d Ising model. We thank the Weizmann Institute of Science, where part of this work was done, and organizers of the ``Back to The Bootstrap 2015'' meeting for hospitality. This work was partially supported by ERC STG grant 306260. L.F.A. is
a Wolfson Royal Society Research Merit Award holder.

\appendix{A}{Examples}

In this appendix we analyze the double light-cone limit of several known perturbative corrections to the four-point correlation functions in different theories. We check that the expected form of the small $u-$ and $v-$ expansion \doubletwistD\ holds. 

The double light-cone expansion of the exact correlators in 2d Ising and Yang-Lee models is presented in appendix B. It takes the form expected from the more general ansatz \ansatzgen . 

\subsec{${\cal N}=4$ SYM}
\noindent {Protected Operators}

\bigskip

The most studied four-point correlator in ${\cal N}=4$ SYM is that of four half-BPS scalar operators of protected dimension 2 that live inside the energy-momentum tensor supermultiplet . These operators transform in the ${\bf 20'}$ representation of the $SU(4)$ R-symmetry group and the correlator has the following structure \EdenBK 
\eqn\largejexpansions{\eqalign{
\langle {\cal O}(y_1,x_1) {\cal O}(y_2,x_2) {\cal O}(y_3,x_3) {\cal O}(y_4,x_4) \rangle = R^{(0)}(y_i,x_i) + R^{(1)}(y_i,y_i) f^{pert}(u,v)
}}
where the harmonic variables $y_i$ encode the $SU(4)$ dependence of the correlator. In the expression above $R^{(0))}(y_i,x_i) $ and $R^{(1))}(y_i,y_i)$ are explicitly known rational functions, see {\it e.g.} \EdenWE. As a consequence of crossing $f^{pert}(u,v)=f^{pert}(v,u)$. At one-loop $f^{pert}(u,v)$ is proportional to the scalar box integral $\Phi(z,\bar z)$. Its standard representation is
%
\eqn\boxA{
\Phi(z,\bar z)= {1 \over z-\bar z} \left(2 {\rm Li}_2(z) - 2 {\rm Li}_2(\bar z) + \log(z \bar z) \log({1-z \over 1-\bar z})  \right)
}
We take the double light-cone limit $u,v \to 0$ such that $z \to 0$ and $1-\bar z \to 0$. Using 
\eqn\ident{
{\rm Li}_2(\bar z)  = -\log(\bar z) \log(1-\bar z) - {\rm Li}_2(1-\bar z)+\zeta_2
}
we obtain a representation for the box function where logarithms and powers in $u,v$ are made explicit and the symmetry between $u$ and $v$ (which corresponds to $z  \leftrightarrow 1-\bar z$) is manifest. More precisely, we obtain\foot{We take the double light-cone limit in two steps. First we expand the box function for small $z,1-\bar z$ and then expand $z,1-\bar{z}$ in powers of $u$ and $v$. This second expansion contains only powers.}
\eqn\boxuv{
\eqalign{
f^{(1-loop)}(u,v) \sim \Phi(u, v)= \sum_{m,n=0} u^m v^n c^{(1)}_{mn}(\log u,\log v) 
}
}
where $c^{(1)}_{mn}(\log u,\log v) $ are polynomials of degree one, with symmetry properties following from $\Phi(u, v)=\Phi(v,u)$. For instance, for the first few orders we obtain
\eqn\boxuvB{
\eqalign{
\Phi(u, v)= &\left( \log u \log v + 2 \zeta_2\right) + \left( \log u \log v +2 \log u +2 \zeta_2 -2 \right) u \cr
&+ \left( \log u \log v +2 \log v +2 \zeta_2 -2 \right) v +\cdots
}
}
The two-loop correlator is given by the following expression \refs{\EdenMV,\BianchiHN}
\eqn\twoloop{\eqalign{
f^{(2-loop)}(u,v) &\sim {1 \over 4(z -\bar z)} \left( \Phi_2(z,\bar z) -\Phi_2(1-z,1-\bar z) - \Phi_2( { z \over z-1},{\bar z \over \bar z-1})  \right) \cr
 &+{2+ 2 z \bar z - z -\bar z \over 16} (\Phi_1(z,\bar z))^2 \ ,
}}
where we have introduced
\eqn\twoloop{
\eqalign{
 \Phi_2(z,\bar z) = 6({\rm Li}_4(z) - {\rm Li}_4(\bar z)) - 3 \log(z \bar z)({\rm Li}_3(z) - {\rm Li}_3(\bar z)) +{1 \over 2} \log^2(z \bar z)({\rm Li}_2(z) - {\rm Li}_2(\bar z)) .
}
}
The small $u,v$ expansion is more cumbersome in this case, but again it can be checked to be of the form \boxuv, where now $c^{(2)}_{mn}(\log u,\log v)$ are polynomials of second order. Hence up to two loop the correlator has the double twist expansion proposed in the body of the paper. Higher order results are available in the literature, see \DrummondNDA, but considering their double light-cone limit expansion beyond the leading term is considerably harder. 

\bigskip

\noindent {Supergravity Result}

\bigskip

The correlator above has also been computed in the planar limit, at large values of the t' Hooft coupling, where it is given by a supergravity approximation. In this limit \refs{\ArutyunovPY ,\DolanTT}
$$f^{(sugra)}(u,v) = -16 u^2 v^2 \bar D_{2422}(u,v)$$
where $\bar D_{\Delta_1,\Delta_2,\Delta_3,\Delta_4}(u,v)$ are the conformal integrals introduced in \refs{\DolanUT,\DHokerPJ} . Among other properties, they satisfy the following crossing identities
\eqn\Dbarident{
\eqalign{
\bar D_{\Delta_1,\Delta_2,\Delta_3,\Delta_4}(u,v) &= v^{-\Delta_2} \bar D_{\Delta_1,\Delta_2,\Delta_4,\Delta_3}(u/v,1/v) \cr
&= \bar D_{\Delta_3,\Delta_2,\Delta_1,\Delta_4}(v,u) \cr
&= u^{-\Delta_2} \bar D_{\Delta_4,\Delta_2,\Delta_3,\Delta_1}(1/u,v/u)
}
}
Which in particular imply $f^{sugra}(u,v)=f^{sugra}(v,u)$. Furthermore, for the case $\Delta_i=1$ we recover the scalar box function

\eqn\Dvsbox{
\eqalign{
\bar D_{1,1,1,1}(u,v) = \Phi(u,v)
}
}
Another useful property is given by the following derivative relations (see for instance \ArutyunovFH)

\eqn\Dder{
\eqalign{
\bar D_{\Delta_1+1,\Delta_2+1,\Delta_3,\Delta_4}(u,v) &= -\partial_u \bar D_{\Delta_1,\Delta_2,\Delta_3,\Delta_4}(u,v) \cr
\bar D_{\Delta_1,\Delta_2+1,\Delta_3+1,\Delta_4}(u,v) &= -\partial_v \bar D_{\Delta_1,\Delta_2,\Delta_3,\Delta_4}(u,v) \cr
\bar D_{\Delta_1,\Delta_2+1,\Delta_3,\Delta_4+1}(u,v) &= (\Delta_2 +u \partial_u +v \partial_v) \bar D_{\Delta_1,\Delta_2,\Delta_3,\Delta_4}(u,v) \cr
}
}
By applying these three relations consecutively we see that $\bar D_{2422}(u,v)$ can be obtained from the scalar box function by applying a differential operator, symmetric in $u$ and $v$. Hence, the double light-cone limit of the supergravity result follows from that of the scalar box function \boxuv and hence has the expected form. 

Many other results available in the literature are given in terms of conformal integrals with integer entries, and it is straightforward to check that these have the proposed double expansion  \doubletwistD. These include the infinite tower of corrections to the supergravity result above constructed in \AldayTSA\ as well instanton corrections to the four point correlator at hand \refs{\BianchiNK , \DoreyPD}. 

\bigskip

\noindent {Unprotected Operators}

\bigskip

As an example of a correlator of four identical unprotected operators in ${\cal N}=4$ SYM, let us consider the correlator of four Konishi operators, of the form ${\cal K} \sim \tr \phi^I \phi^I$. This has been computed to one-loop in \BianchiCM\ with the result

\eqn\Kcorrelator{
\eqalign{
{\cal G}_{\cal K}(u,v)= 1+u^\Delta ({1 \over v^\Delta}+1)+{1 \over 6 c} {\left( u \over v \right)}^{\Delta/2}(1+ u^{\Delta/2}+v^{\Delta/2})+{g^2 N \over 4 \pi^2} {\cal G}^{(1-loop)}_{\cal K}(u,v) + \cdots
}
}
where
\eqn\Koneloop{
\eqalign{
 {\cal G}^{(1-loop)}_{\cal K}(u,v)= &- {u(1+u+v) \over c v} -{u(3-6u+3v) \over 12 c v} \log u -{u(3-6v+3u) \over 12 c v} \log v \cr
 & - {u(1+u^2 +v^2 +4 u+4 v+4 u v) \over 12 c v} \Phi(u,v) .
}
}
The double light-cone limit of this correlator then follows from that of the scalar box function, and has the proposed form. 

\subsec{Critical $O(N)$ $\sigma$-model}

In the following we focus on the correlator of four fundamental spin fields $S_a(x_i)$, $a=1,...,N$ in the critical $O(N)$ $\sigma$-model
\eqn\critonsigma{
\langle S_a(x_1)S_b(x_2)S_c(x_3)S_d(x_4) \rangle = {{\cal G}_{abcd}(u,v) \over {(x_{12}^2x_{34}^2)^{\mu-1}}} \ , 
}
with $\mu=d/2$. In \LangKP\ the ${1 \over N}$-correction to the large $N$ correlators for $2<d<4$ where given. The result is given by the sum of three contributions, denoted by $B_1,B_2$ and $B_3$, proportional to different tensor structures. In our conventions these contributions are given by
\eqn\ONcorrections{
\eqalign{
B_1 &= u^{\mu-1} B(u,v) \,  \cr
B_2 &= {\left( u \over v \right)}^{\mu -1} B({u \over v},{1 \over v}) \ , \cr
B_3 &= B({1 \over u},{v \over u}) \ .
}}
where $B(u,v)$ is given in \LangKP as a series expansion around $u=0$ and $v=1$
\eqn\Bexpansion{
\eqalign{
B(u,v)=& \sum_{m,n=0}^\infty {(m+n)! \over m! n!} { (\mu-1)_{n+m} (\mu-1)_n \over (\mu)_{2n+m}} u^n (1-v)^m \times \left[  -\log u + \psi(n+1)\right.\cr
 & \left.  -\psi(\mu-1+n)-\psi(n+m-1)-\psi(\mu-1+n+m)+2\psi(\mu+2n+m) \right] \ 
}}
up to a normalization factor which is not important for our purposes. Carefully comparing this expansion with the expansion for the ${\bar D}$ functions given in \DolanIY\ we note that
\eqn\BvsDbar{
\eqalign{
B(u,v)={\bar D}_{\mu-1,1,\mu-1,1}(u,v)
}}
Crossing symmetry of the correlator under exchange of operators $1$ and $3$ implies the following combinations should be symmetric under $u \leftrightarrow v$
\eqn\symmetriccombinations{
\eqalign{
v^{\mu-1} B_1 &= u^{\mu-1} v^{\mu-1} {\bar D}_{\mu-1,1,\mu-1,1}(u,v) \cr
v^{\mu-1}(B_2+B_3) &= u^{\mu-1} v {\bar D}_{\mu-1,1,1,\mu-1}(u,v) + u v^{\mu-1}{\bar D}_{1,1,\mu-1,\mu-1}(u,v)\cr
}}
The symmetry of both combinations is indeed guaranteed by the relations \Dbarident. Next, we would like to consider the small $u,v$ expansion of the above combinations. We show in appendix D how to do this systematically. The general structure is of the form

\eqn\smalluvON{
\eqalign{
v^{\mu-1} B_1 &= u^{\mu-1} v^{\mu-1} \sum_{m,n=0}\alpha_{mn}(\log u,\log v) u^m v^n \cr
v^{\mu-1}(B_2+B_3) &= u^{\mu-1} v^{\mu-1} \sum_{m,n=0}\beta_{mn}(\log u,\log v) u^m v^n \cr
& + u^{\mu-1} v \sum_{m,n=0} \gamma_{mn}(\log u,\log v) u^m v^n +v^{\mu-1} u \sum_{m,n=0} \gamma_{mn}(\log v,\log u) v^m u^n
}}
where $\alpha,\beta,\gamma$ are polynomials of first order and $\alpha_{mn}(\log u,\log v)=\alpha_{nm}(\log v,\log u)$ and similar for $\beta$. We hence obtain a double twist expansion corresponding to twists $\tau=d-2+2n$ and $\tau=2n$, which are the twists of the large $N$ theory, dressed with logarithms, in agreement with \doubletwistD.

\appendix{B}{Double Light-cone Limit In The 2d Ising And Yang-Lee Models}

We consider the correlator of four spin fields in the 2d Ising model. The result is \BelavinVU

\eqn\result{\eqalign{
\la \sigma(x_1) \sigma(x_2)  \sigma(x_3) \sigma(x_4) \ra &= {1 \over (x_{13}^2 x_{24}^2)^{{1 \over 8}}} {u (\theta, \bar \theta) \over \left[ u v \right]^{{1 \over 8}}} , \cr
u (\theta, \bar \theta) &= \cos ({ \theta - \bar \theta \over 2} ), \cr
z &= (\sin \theta)^2, ~~~ \bar z = (\sin \bar \theta)^2.
}}

First we notice that the symmetric function that we discussed in the main body of the text is 
\eqn\simplify{
f(u,v) =u (\theta, \bar \theta) = \sqrt{ {1 + \sqrt u + \sqrt v \over 2 }}.
}

As expected it is symmetric $f(u,v) = f(v,u)$ it admits double twist expansion of the form
\eqn\doubletwist{
f(u,v) = \sum_{m,n=0}^{\infty} (-1)^{1-m-n} {\Gamma (m+n - {1 \over 2}) \over \Gamma(m+1) \Gamma(n+1)} {u^{n/2} v^{m/2} \over 2 \sqrt{2 \pi}}.
}

We can rewrite it as sum over residues for $x,y = {\rm integer} \geq 0 $ of the following function
\eqn\doubletwistB{
f(u,v) = -\sum_{{\rm integer \ poles}} \Gamma(-x) \Gamma(-y) \Gamma( x + y - {1 \over 2}){u^{x/2} v^{y/2} \over 2 \sqrt{2 \pi}}.
}

We can try to interpret this as the contour integral along the imaginary axis but we run into problem that the pole at $x+y - {1 \over 2} =0$ contribute. Thus we choose the contour with $-{1 \over 4} < {\rm Re}[x,y]<0$ but such that the pole at $x + y - {1 \over 2} = 0$ is traversed from the right. In this way we recover the Mellin representation of the correlator \joao

\eqn\mellind{
f(u,v) = - \int_{ {\cal C}} {d x d y \over (2 \pi i)^2} \Gamma(-x) \Gamma(-y) \Gamma( x + y - {1 \over 2}){u^{x/2} v^{y/2} \over 2 \sqrt{2 \pi}}.
}

Similarly, we can consider the four-point function in the 2d Yang-Lee model \CardyYY . In this case external operators have dimension $\Delta = - {2 \over 5}$ and the four-point function takes the form
\eqn\fourpoint{
f(u,v) = | _2 F_1 ({3 \over 5}, {4 \over 5}, {6 \over 5},z)|^2 - {\Gamma({6 \over 5})^2 \Gamma({1 \over 5} \Gamma({2 \over 5})\over \Gamma({3 \over 5}) \Gamma({4 \over 5})^3} {1 \over |z|^{2 \over 5}} | _2 F_1 ({3 \over 5}, {2 \over 5}, {4 \over 5},z)|^2 .
}

One can check that again it has the double twist expansion for small $u$ and $v$ of the expected form which can be concisely encapsulated by the following Mellin amplitude
\eqn\mellindB{\eqalign{
f(u,v) &= c_0 \int_{ {\cal C}} {d x d y \over (2 \pi i)^2} \Gamma( x + y + {1 \over 5}) \Gamma( x + y + {2 \over 5}) \Gamma(-x) \Gamma({1 \over 5} - x) \Gamma(-y) \Gamma({1 \over 5} -y) u^{x} v^{y}  \ , \cr
c_0 &= - {2^{{1 \over 5}} \Gamma({11 \over 10}) \Gamma({6 \over 5}) \over \sqrt{5 \pi } \Gamma({9 \over 5})} ,
}}
where the contour goes along the imaginary axis with ${\rm Re}[x,y]<0$ but ${\rm Re}[x + y]>-{1 \over 5}$.

From the point of view of the double light-cone expansion Mellin variables $x$ and $y$ play the role of complexification of the twist in the $u$- and the $v-$ channel correspondingly. Of course, similar complexification of spin is familiar from the Regge theory. 

\appendix{C}{Case Of Fermions And Vectors}

In the case of fermions we can consider, for example, the four-point function of operators ${\cal O} = \bar \psi \psi$. It takes the following form
\eqn\funccrossrat{\eqalign{
{\cal G}(u,v) &= {u^{d-1}+v^{d-1}+u^{d-1} v^{d-1} \over v^{d-1} } \cr
&+{1 \over c} \left( \left({u \over v}\right)^{d/2 - 1} {1 - u - v \over v} + u^{d-1} v^{-d/2} [u-v-1] + u^{d/2 - 1} [v-u-1] \right) .
}}

The small $u$- expansion is given similarly to the case of scalar by the following expression

\eqn\smallufermions{
{\cal G}(u,v) = 1 + {1 \over c} {1-v-v^{d/2} +v^{d/2+1}\over v^{d/2}}
u^{d/2-1}+...}
The relevant three-point OPE coefficients  take the form

\eqn\result{\eqalign{
a_s = 2^{5-d-2 s} \sqrt{\pi} { \Gamma (s+{d \over 2} - 1)  \Gamma (s+ d -2) \over \Gamma(d/2 )^2 \Gamma(s) \Gamma(s+{d-3 \over 2}) }
}}

Similarly, we can consider the theory of free $n$-form field in $d = 2n$ dimensions. In this theory the lowest twist scalar operator is  ${\cal O}  = F^2$. Some basic properties of this theory can be found in \BuchelSK . The result for the four-point function is a sum of three diagrams where a given diagram is a set of contractions of the basic tensors $I_{\mu \nu}(x)$ considered for example in \DolanUT . 

Different contractions come with different signs. Moreover, contractions can form cycles of length $4,8,12,...$. Notice that the body of the cycle (say how we contract points between $1$ and $4$) is the same for all cycles. Thus, we only need to count the way we contract $n$ indices. For example if we contract $i \to i$ we get a contribution to the correlator of the form $I_4^n$ where $I_4={\rm Tr}[I_{12} I_{23} I_{34} I_{41}]$. If we contract indices with some non-trivial braiding for example we contract $2 \to 1$ and $1 \to 2$ and the rest trivially we get three effects. First of all we get $-1$ due to the permutation, second we get $I_8 I_4^{n-2}$ where 
\eqn\traceeight{
I_8={\rm Tr}[I_{12} I_{23} I_{34} I_{41} I_{12} I_{23} I_{34} I_{41} ]
}
and third we get the coefficient in front which is the number of pairs we can choose out of $n$ bins. Proceeding in the analogous manner we get for the $1234$ diagram

\eqn\onetwothreefour{
f_{1234}(u,v) ={1 \over c} \left( {u \over v} \right)^n \left( I_4^n - {n (n-1) \over 2} I_8 I_4^{n-2} + {n (n-1)(n-2)\over 3} I_{12} I_4^{n-3} ... + (-1)^{n-1} \Gamma(n) I_{4 n} \right) .
} 

For example for $n=5$ we get
\eqn\onetwothreefourFIVE{
f_{1234}(u,v) ={1 \over c} \left( {u \over v} \right)^5 \left( I_4^5  - 10 I_8 I_4^3 + 20 I_{12} I_4^2 + 15 I_8^2 I_4 - 30 I_{16} I_4  - 20 I_{12} I_8 + 24 I_{20} \right) .
} 

For general $n$ we get the following result
\eqn\onetwothreefourRes{
f_{1234}(u,v) ={1 \over c} \left( {u \over v} \right)^n \left( { u^2 + (1 - v)^2 + 2 u ({n-1 \over n} v - 1) \over u v } \right) .
}

Thus, the result for the full correlator is given by
\eqn\fullcorr{\eqalign{
G(u,v) &= 1 + u^{2 n} + \left( {u \over v} \right)^{2 n} \cr
&+{1 \over c} \left( {u \over v} \right)^n \left( { (1-u-v)^2 - {2 \over n} u v \over u v } \right) \cr
&+{1 \over c}  \left( {u \over v} \right)^n u^n \left( {(1+v-u)^2 - {2 \over n } v \over v } \right) \cr
&+ {1 \over c}  u^{n} { (1+u-v)^2 - {2 \over n} u \over u}
}}

The relevant collinear asymptotic is
\eqn\collvector{
G(z, \bar z) = z^{{d-2 \over 2}} {1 \over c} \left( {\bar z \over 1 - \bar z } \right)^{{d \over 2} + 1}  \left( 1 + (1 - \bar z)^{{d \over 2} + 1} \right) = z^{{d-2 \over 2}} f_{coll} (\bar z)  .
}

Notice that it is identical to the scalar one if we shift $d \to d' - 4$. Thus, we have
\eqn\expansionvtwo{\eqalign{
f_{coll} (\bar z) &={2 \over c} \sum_{s} a_s \ \bar z^{{d'-2 \over 2}+s} \ _2 F_{1} ({d'-2 \over 2}+ s, {d'-2 \over 2}+s, d'-2 + 2 s , \bar z), \cr
a_s &= {(1 + (-1)^s ) \over 2} 2^{4-d'-2 s} \sqrt{\pi} { \Gamma (s+{d' \over 2} - 1)  \Gamma (s+ d' -3) \over \Gamma(d'/2 - 1)^2 \Gamma(s+1) \Gamma(s+{d'-3 \over 2}) } .
}}

The collinear block $(d',s)$ is equal to the collinear block $(d,s+2)$. Thus, we get
\eqn\expansionvthree{\eqalign{
f_{coll} (\bar z) &={2 \over c} \sum_{s=2}^{\infty} a'_{s} \ \bar z^{{d-2 \over 2}+s} \ _2 F_{1} ({d-2 \over 2}+ s, {d-2 \over 2}+s, d-2 + 2 s , \bar z), \cr
a'_s &={(1 + (-1)^s ) \over 2}  2^{4-d-2 s} \sqrt{\pi} { \Gamma (s+{d \over 2} - 1)  \Gamma (s+ d -1) \over \Gamma(d/2 + 1)^2 \Gamma(s-1) \Gamma(s+{d-3 \over 2}) } .
}}

\appendix{D}{Double Light-Cone Limit Of Conformal Integrals}

In this appendix we show how to systematically compute the small $u,v$ limit of the conformal functions $\bar D_{\Delta_1,\Delta_2,\Delta_3,\Delta_4}(u,v)$. In perturbative computations for conformal field theories often one finds $\bar D$-functions evaluated in other regions, but one can use the identities \Dbarident\ to use the method below. We start with the expansion given in \DolanIY

\eqn\Dbarexp{
\eqalign{
\bar D_{\Delta_1,\Delta_2,\Delta_3,\Delta_4}(u,v) =&\Gamma(-s) {\Gamma(\Delta_1) \Gamma(\Delta_2)\Gamma(\Delta_3+s) \Gamma(\Delta_4+s) \over \Gamma(\Delta_1+\Delta_2)} \cr
& \times G(\Delta_2,\Delta_3+s,1_s,\Delta_1+\Delta_2;u,1-v)\cr
&+\Gamma(s) {\Gamma(\Delta_1-s) \Gamma(\Delta_2-s)\Gamma(\Delta_3) \Gamma(\Delta_4) \over \Gamma(\Delta_3+\Delta_4)}\cr
& u^{-s} G(\Delta_2-s,\Delta_3,1-s,\Delta_3+\Delta_4;u,1-v)
}
}
where $s=1/2(\Delta_1+\Delta_2-\Delta_3-\Delta_4)$ and 
\eqn\Gdef{
\eqalign{
G(\alpha,\beta,\gamma,\delta; u,1-v) = \sum_{m,n=0}^\infty {(\delta-\alpha)_m(\delta-\beta)_m \over m! (\gamma)_m} {(\alpha)_{m+n}(\beta)_{m+n}\over n! (\delta)_{2m+n}} u^m(1-v)^n
}
}
Now for each fixed power of $u$, we perform the sum over powers of $1-v$ to obtain 

\eqn\Gnew{
\eqalign{
G(\alpha,\beta,\gamma,\delta; u,1-v) =& \sum_{m=0}^\infty  {(\delta-\alpha)_m(\delta-\beta)_m \over m! (\gamma)_m} {(\alpha)_m (\beta)_m \over (\delta)_{2m}}\cr
&\times ~_2F_1(\alpha+m,\beta+m,\delta+2m;1-v) u^m
}
}
Now with the use of the identity

\eqn\hyperidentity{
\eqalign{
~_2F_1(a,b,c;z)= &{\Gamma(c)\Gamma(a+b-c) \over \Gamma(a) \Gamma(c)}(1-z)^{c-a-b}~_2F_1(c-a,c-b,c-a-b+1;1-z)\cr
& + {\Gamma(c)\Gamma(c-a-b) \over \Gamma(c-a)\Gamma(c-b)} ~_2F_1(a,b,a+b-c+1;1-z)
}
}
we obtain an expression that can be easily expanded for small $u,v$. In many applications, the parameters $\Delta_i$ take integer values where some terms above contain poles. In all the cases we have analyzed one can consider a limit, where one deforms some of the parameters and then takes away the deformation.

\listrefs

\bye